\documentclass[pre,twocolumn,superscriptaddress,preprintnumbers,aps,showpacs,amsmath,amssymb]{revtex4}

\usepackage[pdftex]{graphicx}
\usepackage{dcolumn}
\usepackage{bm}
\usepackage{color}
\usepackage{longtable}
\usepackage{xr}
\newcommand{\pdiff}[2]{\frac{\partial #1}{\partial #2}}
\newcommand{\odiff}[2]{\frac{d #1}{d #2}}
\newcommand{\cosT}{\cos \theta}
\newcommand{\sinT}{\sin \theta}

\newcommand{\NY}[1]{{\color{black} #1}}
\allowdisplaybreaks[3]

\usepackage{amsmath}	
\begin{document}

\preprint{}

\title{
Spontaneous motion and deformation of a self-propelled droplet 
}

\author{Natsuhiko Yoshinaga }
\email[E-mail: ]{yoshinaga@wpi-aimr.tohoku.ac.jp}
\affiliation{
WPI - Advanced Institute for Materials Research, Tohoku University,
Sendai 980-8577, Japan
}




\begin{abstract}
The time evolution equation of motion and shape are derived for a
 self-propelled droplet driven by a chemical reaction.
The coupling between the chemical reaction and motion makes an
 inhomogeneous concentration distribution as well as a surrounding flow leading to the instability of
 a stationary state.
The instability results in spontaneous motion by which the shape of the
 droplet deforms from a sphere. 
We found that the self-propelled droplet is elongated perpendicular to
 the direction of motion and is characterized as a pusher.
\end{abstract}

\pacs{82.40.Ck, 62.20.F-, 47.63.-b}

\maketitle

\section{Introduction}

Active matters are assemblages of moving elements individually fueled by
energy source \cite{ramaswamy:2010}.
The studies in this field include cell motility
\cite{BernheimGroswasser:2002,gucht:2005,gerbal:2000} and moving droplet
\cite{dosSantos:1995,Toyota:2009,sumino:2005,nagai:2005,Thutupalli2011,thakur:2006}
for an individual level, and also include collective motion of fishes
and flocks of birds \cite{Vicsek:2012}.
Such spontaneous motion is not driven by external force but is sustained
under a force-free condition. 
\NY{
This requires breaking
translational invariance in space or creating an irreversible cycle in time.
}
The broken symmetry is either extrinsic that is imposed externally by
boundary conditions or by material properties\cite{anderson:1989,howse:2007,jiang:2009}  or intrinsic namely nonlinear coupling makes an
isotropic state unstable
\cite{Yoshinaga:2012a,Yabunaka:2012,Schmitt:2013,Michelin:2012}.
The latter mechanism makes the system going to lower symmetry.
When there is relative distance between
a propelled object and another component, the translational symmetry is
broken and a steady velocity emerges \cite{krischer:1994,Shitara:2011}.


The words {\it pusher} and {\it puller} are sometimes used in order to
characterize properties of active matters \cite{hatwalne:2004,Baskaran2009}.
Each active element (swimmer) either pushes or pulls surrounding fluid,
and creates a force dipole with a sign depending on whether the element is
pusher or puller.
\NY{
The surround flow plays an important role in the interaction between swimmers, and
interaction between a swimmer and a wall (see Table.~\ref{table.pusher}
and Sec.~\ref{section.pusher}
for comparison between pusher and puller).
}
Besides the level of individual and a few swimmers, it has been argued
that the sign of force dipole is associated with macroscopic viscosity of
suspension consisting from swimmers \cite{hatwalne:2004,Rafai:2010,Ryan:2011}.
In continuum description, force dipole is closely related to active
stress added to conventional Navier-Stokes and nematohydrodynamic equations.
\NY{
Consequently, whether a swimmer is pusher or puller, namely the sign of
a force dipole plays an important role in determining qualitative
patterns of hydrodynamic instabilities \cite{Edwards:2009}.
}
Such hydrodynamic equations are derived from conservation laws and
symmetry arguments, and analyzed with phenomenological coupling
constants \cite{kruse:2005,Toner:2012}.
The interpretation of the coupling constants from the properties of
individual active elements and their interactions is also discussed for
a few particular systems \cite{aranson:2006,Leoni:2010a,Peshkov:2012b}. 
It is of relevance to analyze another active system and discuss
interpretations of coupling constants appear in possible hydrodynamic equations. 

A spontaneously moving object does not necessarily preserves its shape.
In fact, a cell of an isotropic shape starts to move by breaking a
symmetry of internal states such as distribution of functional proteins
\cite{verkhovsky:1999,yam:2007}.
Although biological systems involve a lot of complex molecules, it has
suggested that there is a relationship between shape and motility of a cell \cite{li:2008,maeda:2008}.
Here in order to ask ourselves whether there is a generic relation
between shape and motility of self-propelled objects, we will discuss a rather simpler system of artificial chemical
molecules with a droplet rather than a model of a cell.

Motion in the absence of an external mechanical force 
has been discussed in terms of the Marangoni effect in which
a liquid droplet is driven by a surface tension gradient
\cite{Young:1959,levan:1981}.
The non-uniform surface tension can be controlled by a field variable
such as temperature and a chemical (typically surfactants) concentration
\cite{darhuber:2005}.
The mechanism is that the gradient induces a convective flow inside and outside
of a droplet, which leads to motion of the droplet itself.
A similar flow and resulting motion are observed for a solid particle in
phoretic phenomena such as thermophoresis \cite{anderson:1989,jiang:2009}.
In both systems, objects are {\it swimming} in a fluid.

\NY{
Even in the absence of such external asymmetry, 
spontaneous motion (and possibly deformation) is realized under a nonequilibrium
state when the droplet has chemical reaction and more specifically is able to
produce chemical molecules from inside.
}
The coupling between the chemical reaction and motion spontaneously
breaks symmetry leading to directional motion.
Surface tension plays an essential role in this system; it serves as a
chemomechanical transducer by which an
inhomogeneous concentration of chemical molecules becomes a mechanical
force acting on a surface of the droplet.

In the present work, 
we derive amplitude equations of motion and shape of a droplet starting from a set of
equations of concentration fields taking hydrodynamics into
consideration.
Spontaneous motion driven by such a chemical reaction was first
discussed by \cite{ryazantsev:1985}.
The present work is an extension of the work \cite{Yabunaka:2012} in
which only a translational motion was discussed.
\NY{
We use a similar method but include deformation and higher moments of a
flow field. 
}
As far as we know, deformation of a spontaneous moving droplet was first
proposed in \cite{golovin:1989}.
Recently a set of amplitude equations of motion and shape has been
phenomenologically proposed also in \cite{ohta:2009b} and is also derived
from reaction-diffusion equations \cite{ohta:2009,Shitara:2011}.
Due to the coupling between the motion and shape, they reproduce complex
self-propulsive motion: not only straight motion but also rotation and
helical motion \cite{Hiraiwa:2011}.
The amplitude equations that we will see in this paper is qualitatively
the same as in \cite{ohta:2009,Shitara:2011} except the coupling constants.
\NY{
Nevertheless, it is worth to mention that a mechanical point of view is
missing in the previous studies and it is not clear how force, more precisely, force moments acting around the
self-propelled particles is associated with motion and deformation.

For a biological cell, a geometric position of the body and propelling objects is associated with its
sign of a force dipole.
E. coli is, for instance, characterized as a pusher; it pushes a fluid back by flagella and pushes forward by
its body. 
In contrast, it is not clear whether a swimmer driven by chemical
reaction is pusher or puller because it does not push or pull
a fluid in an apparent way. 
It is not even clear whether it has a force dipole.
This is important since far-field flow is dominated by a force dipole,
which play a dominant role in an interaction between swimmers.
The previous phenomenological models do not answer to these points.
The purpose of this paper is to clarify physical meanings of coupling
constants appeared in the amplitude equations and to get insight about
these points.
We will show a clear evidence that the chemical swimmer indeed has a force dipole and the sign
is characterized as a pusher.
We will also obtain an explicit form of an active stress created by a self-propelled droplet
using chemical reactions.
}

This paper is organized as follows. 
In Sec.~\ref{section.pusher}, we discuss force moments
under an inhomogeneous surface tension.
In Sec.~\ref{sec.motion}, we present a
model of a deformable droplet under chemical reaction. 
In Sec.~\ref{sec.concentration}, we discuss expansion of a concentration
field around the critical point of nonequilibrium phase transition
between stationary state and self-propelled state.
In Sec.~\ref{sec.deformation}, we derive time evolution equations of deformation.
By considering
flow and concentration fields, we are led to introduce traceless
symmetric tensors associated with a shape of the droplet.
Sec.~\ref{section.amplitude.eq} is a main result; we show amplitude
equations.
The relation to the mechanical structure in Sec.~\ref{section.pusher} is
explained by showing the coefficients appeared in the equations.
\NY{
We conclude with Sec.~\ref{sec.discussions} and Sec.~\ref{sec.summary} by summarizing our results and discussing
their relevance to other studies of self-propelled particle with
deformation.
There is technical overlap between the present work and the previous
works in
\cite{Yabunaka:2012} and \cite{Shitara:2011}.
 Those who are familiar with these works may go  directly from
 Sec.~\ref{section.pusher} to Sec.~\ref{sec.deformation} to
 find essential physics in this work avoiding technical developments.
}
The details of calculation are summarized in Appendix and Supplemental
Material \cite{noteSM}. 
Readers who are
interested in the techniques used in this paper may consult them. 

\section{pusher or puller}
\label{section.pusher}

In this section, we discuss a flow field and mechanical structures under
an inhomogeneous surface tension.
To do this, we perform multipole expansion, namely expand force acting on a droplet into its moment.
As we will see, motion driven by a gradient of surface tension is
force-free \cite{Young:1959,levan:1981}.
This gives a vanishing first moment.
\NY{
The velocity of a droplet and a surrounding flow field obtained in this
section have already been shown by solving a boundary value problem for example in \cite{levan:1981}.
Nevertheless, it is worth to see these results in different
perspectives, namely in terms of force multipoles.
}
We discuss higher moments and clarify the relation between a force dipole
and a second mode of surface tension distribution as well as the relation
between the velocity of a droplet and a third mode of the distribution.

Pushers and pullers are associated with a sign of force dipole \cite{hatwalne:2004}.
The simplest model is two connected beads under anti-parallel force
acting on them.
If the system is totally symmetric, that is, two beads are identical,
the swimmer does not move and is called shaker \cite{Baskaran2009}.
Self-propulsion is realized either by asymmetry in space (such as
difference in the frictional
force arising from different size of beads) and asymmetry in time (irreversible
cycle of motion of beads)\cite{najafi:2004}.
\begin{table*}
\begin{ruledtabular}
\begin{tabular}{c c c}
\hline
& pusher & puller \\
\hline \hline
force dipole & $\leftarrow$ $\rightarrow$ &  $\rightarrow$ $\leftarrow$  \\
viscosity\cite{hatwalne:2004,Haines2009} & reduction & increase \\
interaction between sides \cite{Lauga2009} & attractive  & repulsive \\
wall-induced rotation \cite{Lauga2009} & parallel to a wall &  perpendicular\\
cell type \cite{Haines2009,Rafai:2010} &spermatozoa, bacteria such as E. coli. 
& alga	 Chlamydomonas \\
surface tension ($l=2$ mode) & $\gamma_2 > 0$ & $\gamma_2 <0$ \\
shape & elongate perpendicular to a direction of motion  & 
elongate parallel to it\\
\hline 
\end{tabular}
\end{ruledtabular}
\caption{The properties of pusher and puller.
The force dipole is aligned in a lateral direction and is described
 using two arrows showing a direction of force.
\label{table.pusher}
}
\end{table*}
The asymmetry is characterized either by extensile force dipole (pusher)
or contractile force dipole (puller).
These force dipoles create a surrounding flow, and therefore play an
important role in the
interaction between self-propelled particles and rheological properties
(summarized in Table.~\ref{table.pusher}).
For example, viscosity increases for puller and decreases for pusher.
This argument is valid only for a low concentration of
self-propelled particles.
When the concentration is high, the tendency can reverse \cite{Ryan:2011}.

Under given surface tension, the force acting on the interface of a
spherical droplet with a size $R_0$ is
\begin{align}
{\bf f}_s
&=
 \kappa  \gamma (\theta,\varphi) {\bf n}
+ 
\nabla_s \gamma (\theta,\varphi) {\bf t}
\nonumber \\
&=
- \frac{2}{R_0} \gamma (\theta,\varphi) {\bf n}
+ \frac{1}{R_0} \pdiff{\gamma (\theta,\varphi)}{\theta}  {\bf t}
+ \frac{1}{R_0 \sinT} \pdiff{\gamma (\theta, \varphi)}{\varphi}
{\bf b}
 \label{force.interface}
\end{align}
where ${\bf n} (\theta, \varphi)$ is the unit normal vector on a
sphere 
in a polar coordinate
$
(x,y,z)=
(r \sinT \cos \varphi,
r \sinT \sin \varphi ,
r \cosT
)
$.
The curvature $\kappa$ on a sphere with radius $R_0$ is $\kappa = -2/R_0$.
$\nabla_s = \frac{1}{R_0} \pdiff{}{\theta} + \frac{1}{R_0
\sinT}\pdiff{}{\varphi} $ is a surface gradient derivative.
The first term in (\ref{force.interface}) describes Laplace pressure and
the other terms show tangential force due to a gradient of surface tension.
The normal vector is also expressed with spherical harmonics of $l=1$ \cite{Arfken:1968}
(see also Appendix~\ref{app.spherical.harmonics}):
\begin{align}
{\bf n}
&=
\left(
\sinT \cos \varphi,
\sinT \sin \varphi,
\cosT
\right)
\nonumber \\
&=
\left(
\sqrt{\frac{2\pi}{3}} (-Y_1^1 + Y_1^{-1})
,i\sqrt{\frac{2\pi}{3}} (Y_1^1 + Y_1^{-1})
, \sqrt{\frac{4\pi}{3}} Y_1^0
\right)
\label{normal.vector}
\end{align}
We also define two unit tangent vectors on a sphere
\NY{
$
{\bf t}
=\left(
\cosT \cos \varphi,
\cosT \sin \varphi,
-\sinT
\right)
,
{\bf b}
=
\left(
- \sin \varphi,
\cos \varphi,
0
\right)
$
}
which satisfy ${\bf n} \cdot {\bf t}={\bf n}\cdot{\bf b}={\bf t}\cdot{\bf
b} =0$.
An arbitrary distribution of surface tension is expressed using
spherical harmonics as
\begin{align}
\gamma (\theta,\varphi) 
&=
\sum_{l,m}
\gamma_{l,m}
Y_l^m (\theta,\varphi).
\label{surface.tension.spherical.harmonics}
\end{align}

\begin{figure}[h]
\begin{center}
\includegraphics[width=0.45\textwidth]{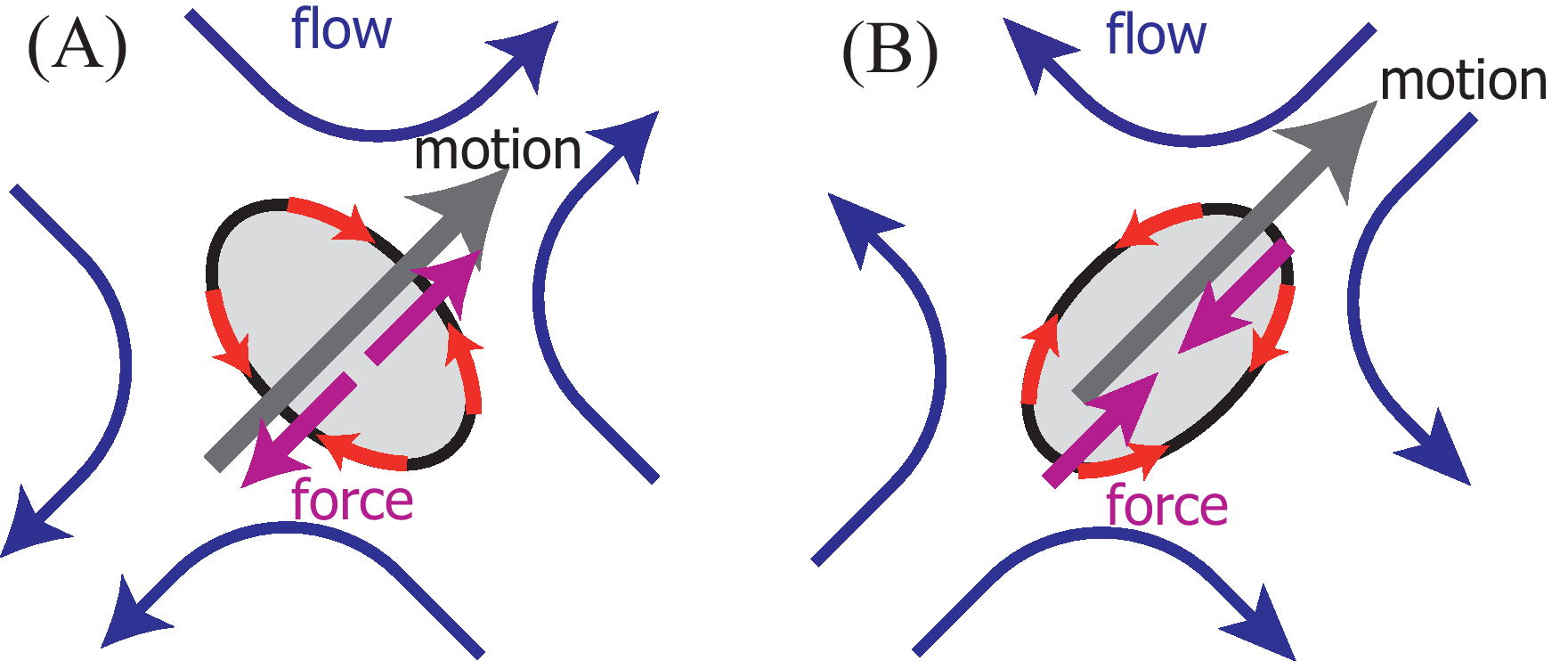}
\caption{ (Color Online) Motion and deformation of {\it pusher} (A) and  {\it puller}
 (B). Under given distribution of the surface tension, the flow field,
 the direction of force dipole,
 and the force acting on the surface of a droplet are drawn depending on
 $\gamma_2 >0$ (A) and $\gamma_2 < 0$ (B).
\label{fig.pusher}
}
\end{center}
\end{figure}

The velocity field created by the force is obtained by solving the
inhomogeneous Stokes equation\cite{Yabunaka:2012}
\begin{align}
\eta \nabla^2 {\bf v} 
- \nabla p
&=- {\bf f}_s \delta (r -R).
\label{Stokes.force}
\end{align}
We apply multipole expansion of the force up to the third moment as
\NY{
\begin{align}
F^{(1)}_j 
&=
\int f_j dV 
\nonumber  \\
F^{(2)}_{jk} 
&=
 \int f_j x_k dV
\label{force.moment}
\\
F_{jkl}^{(3)} 
&= 
 \int f_j x_k x_l dV 
\nonumber
.
\end{align}
}
The velocity field is expanded as \cite{Kim:1991}
\begin{align}
v_i(x) 
&=
 \mathsf{T}_{ij} F_j^{(1)}
- \mathsf{T}_{ij,k} F_{jk}^{(2)}
+ \frac{1}{2} \mathsf{T}_{ij,kl} F_{jkl}^{(3)}
- \cdots
\label{multipolev.3D} 
.
\end{align}
Throughout the paper, the convention of summation over repeated indices applies.
The Oseen tensor $\mathsf{T}_{ij}$ is
\begin{align}
\mathsf{T}_{ij}  
&=
\frac{1}{8\pi \eta}
\left[
\frac{1}{r} \delta_{ij}
+ \frac{x_i x_j}{r^3} 
\right].
\end{align}
The moments are explicitly shown here when we choose an appropriate axis
i.e., by taking $\gamma_{lm}=0$ for $m\neq 0$.
\begin{align}
F_{jk}^{(2)}
 &=
\frac{4 \pi R_0^2 \gamma_2}{15}
\begin{pmatrix}
-1  & 0 & 0 \\
0 & -1 & 0\\
0 & 0 & 2
\end{pmatrix}
\end{align}
and
\begin{align}
&
 F_{jkl}^{(3)} 
=
\frac{4 \sqrt{\pi} R_0^3}{5}
\nonumber \\
&\times
\begin{cases}
 \frac{1}{\sqrt{3}} \gamma_1 - \frac{1}{\sqrt{7}} \gamma_3
&\mbox{   for  } (j,k,l) =(3,1,1),(3,2,2) 
\nonumber \\
 - \frac{2}{\sqrt{3}} \gamma_1 + \frac{2}{\sqrt{7}} \gamma_3
&\mbox{   for  } (j,k,l) =(3,3,3) 
\nonumber \\
- \frac{\sqrt{3}}{2} \gamma_1 - \frac{1}{\sqrt{7}} \gamma_3
&\mbox{    for  } (j,k,l)=(1,1,3),(2,2,3),
\nonumber \\
&\mbox{         }
\qquad \qquad \qquad (1,3,1),(2,3,2) 
\nonumber \\
0 
&\mbox{   otherwise   }
\end{cases}
\end{align}
where $\gamma_2$ terms vanish in $F^{(3)}_{jkl}$.
Since there is no force acting on the system, the first moment $F^{(1)}_j$
is zero as it should be.
The second moment describes $l=2$-mode deformation.
When $\gamma_2 > 0$, the direction of the force (stress) is expanding in $z$-direction and
contracting in $x$- and $y$-direction. 
On the other hand, the direction of the force is contracting in $z$-direction and
expanding in $x$- and $y$-direction for $\gamma_2 < 0$.
Therefore, if we choose $z$-axis as a direction of motion, $\gamma_2>0$
 corresponds to pusher and $\gamma_2 < 0$ corresponds to puller
\cite{Schmitt:2013}.
If we carefully have a look at the two terms in (\ref{force.interface}), the second terms dominate the forces.
The forces arising from the Laplace pressure (the first terms) have
opposite directions of forces, namely contracting in $z$-direction for
$\gamma_2 > 0$.
However the force from surface tension gradient has a stronger
contribution due to the factor $n$ appearing from derivative with
respect to $\theta$.
A shape of the droplet, on the other hand, is determined by the Laplace
pressure since it is associated with a normal force.
When $\gamma_2 > 0$, the Laplace pressure is larger at the front and
the back along the direction of motion (Fig.~\ref{fig.pusher}(A)).
Therefore, the interface is flatter at higher surface tension (see Fig.~\ref{fig.pusher}).

The velocity field for an arbitrary direction of motion and deformation
is expressed using vector spherical harmonics as \cite{Thorne:1980} (see
also Appendix~\ref{app.spherical.harmonics})
\begin{align}
{\bf v} 
=
\sum_{l=1}^{\infty}
{\bf v}_l
=
\sum_{l=1}^{\infty}
\sum_{m=-l}^{l}
&
\left[
v_{lm}^{r} (r) 
{\bf Y}_l^m (\theta,\varphi)
+ v_{lm}^{\Psi} (r) 
{\bf \Psi}_l^m (\theta,\varphi)
\right.
\nonumber \\
&
\left.
+ v_{lm}^{\Phi} (r) 
{\bf \Phi}_l^m (\theta,\varphi)
\right]
\label{vel.3D.asym.exp}
.
\end{align}
\NY{
The coefficients for $l=1$ is obtained from (\ref{force.moment})
and (\ref{multipolev.3D}) as
\begin{align}
\left(
v_{1m}^{r},
v_{1m}^{\Psi},
v_{1m}^{\Phi}
\right)
&=
\left(
- \frac{2 R^3}{15 \eta r^3}
\gamma_{1,m},
\frac{ R^3}{15 \eta r^3}
\gamma_{1,m},
0
\right).
\label{vel.3D.asym.exp2a}
\end{align}
}
Vanishing the coefficient of ${\bf \Phi}_l^m(\theta,\varphi)$ arises
from that there is no helical force acting on the droplet since the
forces in two radial directions originate from the gradient of
a scalar variable of surface tension.

\NY{
The velocity of the droplet given by (\ref{vel.geometric})
for a spherical droplet with (\ref{vel.3D.asym.exp}) and
(\ref{vel.3D.asym.exp2a}) becomes
}
\begin{align}
{\bf u} 
&=
- \frac{2 }{15 \eta }
\frac{3}{4\pi}
\int 
 {\bf n}(\theta,\varphi)
\gamma_{1,m}
Y_1^m(\theta, \varphi)
\sinT d\theta d\varphi
da 
\nonumber \\
&=
- \frac{2 }{15 \eta }
\sqrt{\frac{3}{4\pi}}
\begin{pmatrix}
\frac{1}{\sqrt{2}}
\left(
-\gamma_{1,1}
+ \gamma_{1,-1}
\right)\\
\frac{i}{\sqrt{2}}
\left(
\gamma_{1,1}
+ \gamma_{1,-1}
\right)
\\
\gamma_{1,0}
\end{pmatrix}
\label{multipole.drop.velocity}
\end{align}
The velocity field and the velocity of the droplet that we obtained is consistent with those obtained from 
a boundary-value problem \cite{Yoshinaga:2012a}.
The origin of the velocity field
(\ref{vel.3D.asym.exp2a}) and the velocity of
the droplet (\ref{multipole.drop.velocity}) is $F_{ijk}^{(3)}$.
Therefore the third moment corresponds to translational motion.
\NY{
This shows that the motion of the droplet is not driven by Stokeslet or
force monopole in (\ref{force.moment}), but by force quadrupole
in (\ref{force.moment}).
}
The same velocity dependence is also observed in another force-free
motion, namely phoretic motion \cite{julicher2009}. 
Note that the coefficients $\gamma_{l,m}$ in
(\ref{surface.tension.spherical.harmonics}) and $\gamma_n$ in
\cite{levan:1981,Yoshinaga:2012a} are different by the factor
$\sqrt{3/(4\pi)}$ due to the definition of spherical harmonics.

The velocity field generated by the force dipole is given by
(\ref{vel.3D.asym.exp}) with coefficients 
\begin{align}
v_{2,m}^r 
&=
- \frac{R^2 }{5 \eta r^2} \gamma_{2,m}
\label{flow.dipole.3D}
\end{align}
and $v_{2,m}^{\Psi} = v_{2,m}^{\Phi} = 0$.
In terms of normal $v_r$ and tangential $v_{\theta}$ velocities,
the velocity field is obtained as $v_{r,l=2} \simeq (R/r)^2$ and
$v_{\theta}=0$.
This is also consistent with the velocity field obtained from 
a boundary-value problem \cite{Yoshinaga:2012a}.
\NY{
As we have seen from (\ref{vel.3D.asym.exp2a}) and (\ref{flow.dipole.3D}),
far-field flow is dominated by the force dipole decaying $1/r^2$ rather
than the force quadrupole $1/r^3$.
}

\NY{
A similar argument is possible in two dimensions.
We should note that there is no Stokes paradox for a droplet moving
under a gradient of surface tension \cite{Kitahata:2013b}.
This is because there is no force monopole in this system and therefore
no logarithmic term as we will see below.
The detail velocity fields are shown in Appendix~\ref{app.2D}.
}

\section{motion of a droplet}
\label{sec.motion}
The surface tension discussed in the previous section is dependent on
concentration field $c({\bf r})$ of chemicals.
A typical example of the chemical is surfactant though our model is
not restricted in the example.
The concentration field is also regarded as a temperature field which
modifies surface tension. 
The surface tension is assumed to be linear in the concentration at the
interface, 
\begin{align}
\gamma (t)
&=
\gamma_0 + \gamma_c c(a,t).
\label{surfacetension.con.lin}
\end{align}
In this paper, we focus on the case of $\gamma_c > 0$.
Yet the argument is straightforwardly extended to another case.

Together with the argument of Sec.~\ref{section.pusher}, the flow field
inside and outside the droplet determined by concentration
at the interface between the droplet and a surrounding fluid.
In this section, we derive a set of kinematic equations showing motion
and deformation of a droplet is determined by the flow field.
The concentration field itself is affected by the position of the
droplet and the flow field as we will discuss in the next section.
These relation forms closed equations of the system.

 \subsection{kinematic equations of a droplet}
\begin{figure}[h]
\begin{center}
\includegraphics[width=0.45\textwidth]{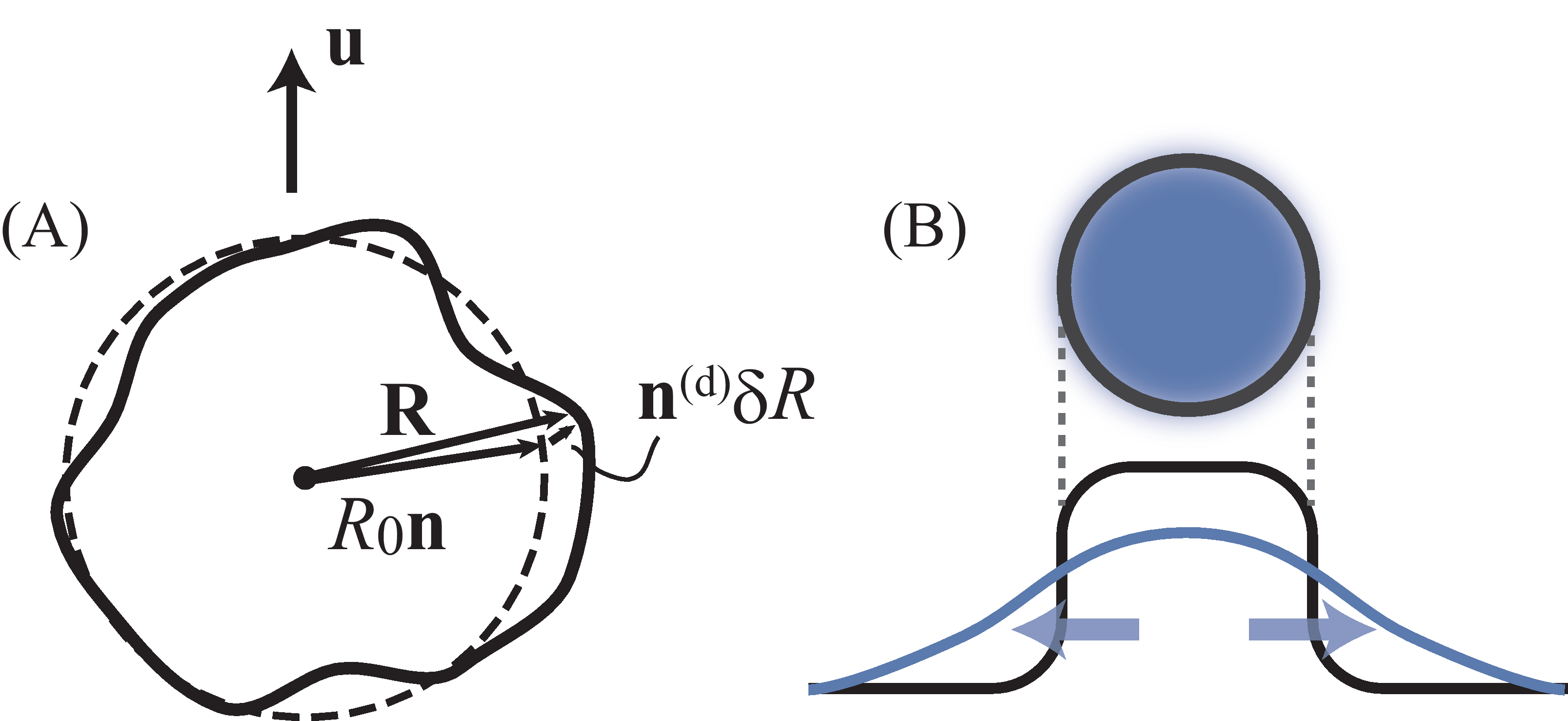}
\caption{(Color Online)
Schematic picture of a self-propelled droplet. 
(A) The deformed interface is drawn by a solid line with respect to a
 reference spherical shape (dashed line).
(B) Concentration distribution across the interface is shown in blue
 (gray) gradation (top) and in a solid blue (gray) line (bottom).
The droplet with a sharp but smooth interface is drawn in a black line. 
\label{fig.deformation}
}
\end{center}
\end{figure}

Let us consider deformation around a spherical droplet.
We will have a kinematic equation of the shape following the step
discussed in \cite{Shitara:2011}.
The surface of the droplet is expressed as
\begin{align}
R(a,t) 
&=
R_0
+ \delta R(a,t),
\end{align}
\NY{
where $R_0$ is an unperturbed radius of a spherical droplet and $\delta
R$ is deviation from it as a function of polar ($\theta$) and azimuthal ($\varphi$) angles
denoted by a surface area $a$ (see Fig.~\ref{fig.deformation}).
}
We may expand the deformation as
\begin{align}
\delta R(\theta,\varphi) 
&=
\sum_{l \geq 2,m} w_{l,m}
Y_l^m (\theta,\varphi)
\label{deformation.wlm}
\end{align}
using spherical harmonics in three dimensions (see Appendix I for the
definition of spherical harmonics with normalization constants).
Note that translational motion is treated independently from deformation
and therefore, $l=1$ is not included in the summation in (\ref{deformation.wlm}). 

The normal velocity $v_n$ and curvature $\kappa$ are defined using a level-set function
\NY{
$
\delta r 
=
r - R(a,t)
$
as\cite{Shitara:2011}
\begin{align}
v_n ({\bf r},t) 
&=
- \frac{1}{|\nabla \delta r|}
\left.
\pdiff{\delta r}{t} 
\right|_{\delta r=0}
\\
\kappa ({\bf r},t) 
&=
- \nabla \cdot 
\left.
\left(
\frac{\nabla \delta r}{|\nabla \delta r|}
\right)
\right|_{\delta r=0}.
\end{align}
}
The unit normal vector on a curved interface is given as a perturbation
around a unit normal vector on a sphere for small
deformation:
\begin{align}
 {\bf n}^{({\rm d})}
&=
\frac{\nabla \delta r}{|\nabla \delta r|}
\simeq
{\bf n} - \nabla_s \delta R\label{unit.vec.def}
.
\end{align}

The position vector on an interface is described at the linear order in
deformation  as (see Fig.~\ref{fig.deformation})
\begin{align}
{\bf R} 
&=
R_0 {\bf n}
+ \delta R {\bf n}^{({\rm d})}
\simeq
{\bf n}
\left(
R_0 + \delta R 
\right)
.
\end{align} 
The normal velocity is expressed as
\NY{
\begin{align}
v_n (a,t) 
&=
{\bf u} \cdot {\bf n}^{({\rm d})}
+ 
\sum_{l,m} \odiff{w_{lm}}{t} 
Y_l^m (a)
\label{vn.def.lin}
\end{align}
and the curvature is approximated for small deformation as
\begin{align}
\kappa (a)
&=
-\frac{2}{R_0}
-\frac{1}{R_0^2}
\sum_{l,m} (l+2)(l-1) w_{lm}(t) Y_l^m(a)\label{surface.tension.def.lin}
.
\end{align}
}
Although the velocity in (\ref{vn.def.lin}) is the velocity of the
contour center of the surface, we will use the velocity of the center of
mass since the two centers are same for a small deformation (see (\ref{vel.geometric})). 

\subsection{velocity of the droplet}
From geometric consideration, the velocity
 of the droplet ${\bf u}$ is given by \cite{Kawasaki1983,noteAvel}
\begin{align}
u_i 
&=
\frac{1}{\Omega}
\int d\tilde{a} v_n(a) R_i(a)
\label{vel.geometric}
\end{align}
where $\Omega$ is the volume of the droplet and is same as that of
undeformed droplet for small deformation
\NY{
$
\Omega 
=
\int R^2 d\tilde{a}
\simeq
\frac{4}{3} \pi R_0^3
.
$
At the first order in deformation,
the infinitesimal area $d\tilde{a}$ on a deformed interface is also
approximated with the undeformed area $da$.
}
The velocity of the droplet (\ref{vel.geometric}) is also rewritten as 
\begin{align}
u_i 
&\simeq
\frac{1}{\Omega}
\int d\tilde{a} v_n(a) 
n_i(a)
\left(
R_0 + \delta R(a)
\right)
\label{self.consist.velocity}
\end{align}

The velocity field driven by the force acting on the interface has two
origins: normal and tangential forces in (\ref{force.interface})
\begin{align}
v_i ({\bf r},t) 
=& 
\int d\tilde{a}'
\mathsf{T}_{ij} ({\bf r},{\bf r}(a'))
n_j^{({\rm d})}(a') \gamma (a',t) \kappa(a',t)
\nonumber \\
&
+
\int d\tilde{a}'
\mathsf{T}_{ij} ({\bf r},{\bf r}(a'))
\mathcal{P}_{jk}^{({\rm d})}(a')
\nabla_k \gamma (a',t) 
.
\label{vel.normal}
\end{align}
The projection of the vector onto the direction perpendicular to
$n_j(a')$
is defined as
\begin{align}
 \mathcal{P}_{jk}^{({\rm d})} (a')
&=
\delta_{jk} 
- n_j^{({\rm d})} (a') n_k^{({\rm d})} (a').
\end{align} 
The first and second terms in (\ref{vel.normal})
originate from normal and tangential forces \cite{Yabunaka:2012}.
The normal velocity is expressed  using (\ref{surfacetension.con.lin})
 as 
\NY{
$
v_n ({\bf r})
=
v_{n,0} ({\bf r})
+
v_{n,1} ({\bf r}) 
+ v_{n,2} ({\bf r})
$
where
\begin{align}
&
\begin{Bmatrix}
v_{n,0} ({\bf r}) \\
v_{n,1}  ({\bf r}) \\
v_{n,2} ({\bf r})
\end{Bmatrix}
\nonumber \\
&=
\int d\tilde{a}'
\mathsf{T}_{ij} ({\bf r},{\bf r}(a'))
n_i^{({\rm d})} (a) 
 \begin{Bmatrix}
\gamma_0
n_j^{({\rm d})} (a') 
 \kappa(a',t)
\\
\gamma_c
 n_j^{({\rm d})} (a') 
c_I (a',t) \kappa(a',t)
\\
\gamma_c
\mathcal{P}_{jk}^{({\rm d})} (a')
\left(
\nabla_k c (a',t)
\right)_I
 \end{Bmatrix}
\label{vel.normal} 
\end{align}
}
where $()_I$ denotes the value taken at the interface.
\NY{
$v_{n,0} ({\bf r})$ describes a flow created during relaxation of a deformed shape, and thus this term
does not make any contributions to the motion of the droplet.
In fact, for a spherical droplet, this term vanishes.
Using (\ref{surface.tension.def.lin}) and (\ref{int.nTnYlmEl}) in \cite{noteSM}, its
surface velocity $v_{n0} ({\bf r}(a))$ is simply expressed by
\begin{align}
v_{n0} ({\bf r}(a))
&\simeq
- \frac{\gamma_0}{R_0^2}
(l+2)(l-1) w_{lm} E_l Y_l^m(a).
\label{vn0}
\end{align}
The velocity of the droplet is also decomposed by two contributions
according to (\ref{vel.normal}) as $u_i =
u_{i,1} + u_{i,2}$.
}
In the next section, we will see that the concentration is also expressed in terms of
velocity of the droplet.
Therefore, (\ref{self.consist.velocity}) is self-consistent equation for
the droplet velocity and indeed in section~\ref{section.amplitude.eq} we
obtain the amplitude equation of velocity from
(\ref{self.consist.velocity}) (see also \cite{Yabunaka:2012}).

The contribution from normal force is further decomposed into 
\begin{align}
u_{i,1}
\simeq &
\frac{\gamma_c}{\Omega}
\int d\tilde{a}
\int d\tilde{a}'
n_i(a)
\left(
R_0 + \delta R(a)
\right)
\mathsf{T}_{jk} ({\bf r},{\bf r}(a'))
\nonumber \\
& \times
n_j^{({\rm d})} (a) n_k^{({\rm d})} (a') 
c (a',t) \kappa(a',t)
\nonumber \\
= &
u_{i,1}^{(0)}
+ u_{i,1}^{(1)}
+ u_{i,1}^{(2)}
+ u_{i,1}^{(3)}
+ u_{i,1}^{(4)}
+ u_{i,1}^{(5)}
\label{velocity.decomp}
,
\end{align}
\NY{
where 
\begin{align}
\begin{Bmatrix}
 u_{i,1}^{(\alpha)}
\end{Bmatrix}
&=
\frac{ \gamma_c }{\Omega}
\int da
\int d a'
n_i(a)
\mathsf{T}_{jk} ({\bf r},{\bf r}(a')
c (a',t)
\nonumber \\
&\times
\begin{Bmatrix}
- 2
n_j (a) n_k(a') 
 \\
- 6
n_j (a) n_k(a') 
 \frac{\delta R(a)}{R_0}
\\
2
 n_k (a') \nabla_{s,j} \delta R(a)
\\
2
n_j (a) \nabla_{s,k} \delta R(a')
\\
-4
 n_j (a) n_k(a') 
 \frac{\delta R(a')}{R_0}
\\
\frac{1}{R_0}
\sum_{l,m} (l+2) (l-1)
w_{lm}(t) Y_l^m (\theta',\varphi')
\end{Bmatrix}
.
\label{ui1}
\end{align}
$u_{i,1}^{(0)}$ describes the distortion of a concentration field and is
same as (22) in \cite{Yabunaka:2012}.
Due to deformation, there are following additional contributions.
}
The deviation of a position at surface from a spherical shape and a
change of a local surface area are given by $u_{i,1}^{(1)}$ and
$u_{i,1}^{(4)}$.
The distortion of a unit vector is included by $u_{i,1}^{(2)}$ and
$u_{i,1}^{(3)}$.
The curvature on a deformed interface is included in $u_{i,1}^{(5)}$.
\NY{
In (\ref{ui1}), 
$u_{i,1}^{(0)}$, $u_{i,1}^{(3)}$ , $u_{i,1}^{(4)}$, and $u_{i,1}^{(5)}$
are immediately simplified using the results in Sec.~\ref{section.Oseen} in \cite{noteSM} for the integral
with Oseen tensor. 
\begin{align}
\begin{Bmatrix}
 u_{i,1}^{(0)}
\\
 u_{i,1}^{(3)}
\\
 u_{i,1}^{(4)}
\\
 u_{i,1}^{(5)}
\end{Bmatrix}
&=
-\frac{2 \gamma_c }{15 \Omega \eta}
\int da' 
\begin{Bmatrix}
4 R_0 n_i (a')
 \\
-3 R_0
 \nabla_{s,i} \delta R(a')
\\
8
 n_i(a') 
 \delta R(a')
\\
2 n_i(a')
\sum_{l,m} (l+2) (l-1)
\\
\times
w_{lm}(t) Y_l^m (\theta',\varphi')
\end{Bmatrix}
 c(a').
\label{ui1.simp}
\end{align}
}
Similarly, the contribution $u_{i,2}$ from tangential force is
\begin{align}
u_{i,2}
\simeq &
\frac{\gamma_c}{\Omega}
\int d\tilde{a}
\int d\tilde{a}'
n_i(a)
\left(
R_0 + \delta R(a)
\right)
\mathsf{T}_{jk} ({\bf r},{\bf r}(a'))
n_j^{({\rm d})} (a)
\nonumber \\
& \times
 \mathcal{P}_{kl}^{({\rm d})} (a')
\nabla_l c (a',t) 
\nonumber \\
=&
u_{i,2}^{(0)}
+ u_{i,2}^{(1)}
+ u_{i,2}^{(2)}
+ u_{i,2}^{(3)}
+ u_{i,2}^{(4)}.
\label{velocity.decomp2}
\end{align}
The detail expressions are shown in Sec.~\ref{app.coeff} in \cite{noteSM}.

\section{concentration field}
\label{sec.concentration}

We have seen a flow and motion of a droplet under a given concentration
field. 
In this section, we consider simple as possible dynamics of the concentration field
making the droplet moving under nonequilibrium states. 
\NY{
Our model is motivated particularly by the experiments of
\cite{Toyota:2009,Thutupalli2011} in which spontaneous motion is
realized with the aid of chemical reaction.
In both experiments, the system is away from equilibrium in the sense
that a droplet either consume or produce molecules that
modify a surface tension.
}
We consider a following reaction-diffusion equation with an internal source of the
chemical,
\begin{align}
\pdiff{c}{t} 
+ {\bf v} \cdot \nabla c
&=
D\nabla^2 c - k (c-c_{\infty})
+A \Theta \left(
R - |{\bf r} - {\bf r}_G|
\right)
\label{con1}
\end{align}
where ${\bf r}_G$ is a center of mass of the droplet.
The same model was analyzed in \cite{Yabunaka:2012} and it was shown
that it exhibits a spontaneous translation motion.
The last term in the right-hand side represents the source with a
magnitude $A$.
\NY{
$A>0$ corresponds to production of molecules  $C({\bf r})$ while $A<0$
corresponds to consumption.
The term of $k$ expresses buffering of a concentration field \cite{Bonnecaze:1991}.
}
Note that we are treating the system in the  laboratory frame, and thus
a fluid flow generated by a gradient of a surface tension on the droplet is
included in two terms.
One is the source term through a position of the droplet ${\bf r}_G$,
which is determined by the fluid flow as in (\ref{vel.geometric}).
Second one is convection and is shown by the second term in the left-hand side of
(\ref{con1}).
The source makes a system apart from an equilibrium state; without this
term no spontaneous motion is realized.
In this sense, this term is essential in our study.
The convective term could be included perturbatively in our result.
However, it is less dominant and it does not modify our results qualitatively \cite{Yoshinaga:2012a,Yabunaka:2012}.
We will therefore neglect the convective term in (\ref{con1}).

In the Fourier space, the equation reads \cite{noteFourier}
\begin{align}
\pdiff{c_{\bf q}}{t}
&=
-D (q^2 + \beta^2) c_{\bf q}
+ H_{\bf q},
\label{con.fourier}
\end{align}
\NY{
with inverse length 
$
\beta 
=
\sqrt{{k}/{D}}.
$
}
The source term is expressed in Fourier space as
\begin{align}
H_{\bf q} 
&=
A S_q e^{i {\bf q} \cdot {\bf r}_G}
\end{align}
with
\begin{align}
S_q 
=&
\int d^3 {\bf r} e^{i {\bf q}\cdot {\bf r}}
\Theta (|{\bf r}| -R)
\nonumber \\
=&
\int_{0}^{R_0} r^2 dr \int _{0}^{\pi}\sinT
\int d\varphi
  e^{i {\bf q}\cdot {\bf r}} 
\nonumber \\
& +
\int_{R_0}^{R_0+ \delta R} r^2 dr \int _{0}^{\pi}\sinT
\int d\varphi
  e^{i {\bf q}\cdot {\bf r}} 
 \nonumber \\
=&
S_q^{(0)}+ S_q^{(1)}.
\label{Sq}
\end{align}
Since only $l=0$ contributes to the first term
, we obtain
\begin{align}
S^{(0)}_q 
&=
4 \pi \frac{\sin (qR) - q R \cos (qR)}{q^3}
= \frac{4\pi R_0^2}{q} j_1(q R_0)
.
\end{align}
where we have used 
\begin{align}
e^{\pm i{\bf q}\cdot {\bf r}} 
&=
4\pi 
\sum_{l,m}
(\pm i)^l j_l (qr)
Y_l^{m*}(\theta_q,\varphi_q) 
Y_l ^m (\theta,\varphi)
\nonumber \\
&= 
4\pi 
\sum_{l,m}
(\pm i)^l j_l (qr)
Y_l^{m}(\theta_q,\varphi_q) 
Y_l ^{m*} (\theta,\varphi)
.\label{eiqrYlm}
\end{align}
Here, $j_l(x)$ is the spherical Bessel function of the first kind \cite{Arfken:1968}.
The second term in (\ref{Sq}) is
\begin{align}
S_q^{(1)} 
\simeq &
4 \pi R_0^2
\int_{0}^{\pi} \sinT \int_0^{2\pi} d\varphi
 \delta R(\theta,\varphi)
\sum_{l,m}
i^l j_l (qR_0)
\nonumber \\
& \times
Y_l^{m*}(\theta_q,\varphi_q) 
Y_l ^m (\theta,\varphi)
\nonumber \\
= &
4 \pi R_0^2
\sum_{l,m}
i^l j_l (qR_0)
w_{lm}^{*}
Y_l^{m*}(\theta_q,\varphi_q).
\end{align}

Following \cite{Yabunaka:2012}, the solution of (\ref{con.fourier}) is expanded close to the critical
point of drift bifurcation, namely for $\epsilon = u/(D\beta) \ll 1$,
\begin{align}
c_{\bf q} 
&=
\frac{G_q}{D} H_{\bf q}
- \frac{G_q^2}{D^2} \pdiff{H_{\bf q}}{t}
+ \frac{G_q^3}{D^3} \pdiff{^2 H_{\bf q}}{t^2}
- \frac{G_q^4}{D^4} \pdiff{^3 H_{\bf q}}{t^3}
+ \cdots
\label{con.sol.fourier}
\end{align}
with the Green's function
\begin{align}
G_q 
&=
\frac{1}{q^2 + \beta^2}.
\end{align}
After inverse Fourier transformation, the concentration at the interface
$c_I$ is following the expansion of (\ref{con.sol.fourier})
\begin{align}
c_I 
= &
c_I^{(0)} ({\bf r}_G + {\bf s}) 
+ c_I^{(1)} ({\bf r}_G + {\bf s}) 
+ c_I^{(2)} ({\bf r}_G + {\bf s}) 
\nonumber \\
&
+ c_I^{(3)} ({\bf r}_G + {\bf s})
+ \cdots,
\label{cI}
\end{align}
where
\begin{align}
c_I^{(0)} ({\bf r}_G + {\bf s}) 
&=
\frac{A}{D} \int_{\bf q}
G_q S_q e^{i {\bf q}\cdot {\bf r}_G}
e^{-i {\bf q}\cdot ({\bf r}_G+{\bf s})}
\nonumber \\
&\simeq
\frac{A}{D} 
\left[
Q^{(0)}_1(s)
+ Q^{(1)}_1(\theta,\varphi)
\right],
\label{cI0}
\end{align}
and
\begin{align}
&
 c_I^{(1)} ({\bf r}_G + {\bf s}) 
= 
- \frac{A}{D^2} \int_{\bf q}
(i {\bf q}\cdot {\bf u})
G_q^2 S_q e^{- i{\bf q} \cdot {\bf s}}
\nonumber \\
= & 
u_i \frac{A}{D^2} 
\left[
 n^{(0)}_i 
\left(
\pdiff{Q^{(0)}_2(s)}{s}
+ \pdiff{Q^{(1)}_2(\theta,\varphi)}{s}
\right)
+ \nabla_{s,i}Q^{(1)}_2(\theta,\varphi)
\right].
\label{cI1}
\end{align}
\NY{
A similar formula is also obtained for $c_I^{(2)} ({\bf r}_G + {\bf s}) $
and $c_I^{(3)} ({\bf r}_G + {\bf s}) $.
Due to deformation, the terms with $Q_n^{(1)}$ appear in addition to the
contribution from a spherical droplet $Q_n^{(0)}$ (see (35)-(38) in
\cite{Yabunaka:2012}).
These novel terms are anisotropic and therefore are associated with
coupling between the motion and deformation.
}
We show the detail form of the perturbed concentration field \NY{in \cite{noteSM}}.
Here the velocity of the droplet is given by
\begin{align}
{\bf u} 
&=
\odiff{{\bf r}_G}{t}
\end{align}
and we have defined
\begin{align}
Q_n ({\bf s}) 
&=
\int_{\bf q} G_q^n S_q e^{-i{\bf q}\cdot {\bf s}}
=
Q_n^{(0)}(s)
+ Q_n^{(1)}(\theta,\varphi)
\end{align}
\NY{
according to (\ref{Sq}).
$Q_n^{(0)} (s) = 
\int_{\bf q} G_q^n S^{(0)}_q e^{-i{\bf q}\cdot {\bf s}}
= 
\frac{2R_0^2 }{\pi}
\int_0^{\infty}  dq 
G_q^n q j_1(q R_0)
j_0 (q s) 
$ is isotropic contribution
although $Q_n$ is not necessarily isotropic due to the anisotropy of
the shape, 
$R(\theta, \varphi)$.
This is shown by the second term $Q_n^{(1)}(\theta,\varphi)$,
}
\begin{align}
 Q_n^{(1)} ({\bf s}) 
&=
\int_{\bf q} G_q^n S^{(1)}_q e^{-i{\bf q}\cdot {\bf s}}
=
\sum_{l,m}
\bar{Q}^{(1)}_{n,l} (s)
w^*_{lm} Y_l^{m*} (\theta,\varphi)
,
\label{Qn1}
\end{align}
with
\begin{align}
 \bar{Q}^{(1)}_{n,l} (s)
&=
\frac{2  R_0^2}{\pi}
\int_0^{\infty} dq 
\sum_{l,m}
G_q^n q^2
j_l(qs)
j_{l} (q R_0)
.
\end{align}

\section{deformation}
\label{sec.deformation}

The shape of the droplet is characterized by tensor order parameters.
This was pointed out for a moving droplet in
\cite{Shitara:2011,Hiraiwa:2011} for the second mode in three dimensions
and also in \cite{ohta:2009} for the third mode in two dimensions.
The tensors are symmetric and traceless so that they satisfy rotational
symmetry and \NY{volume conservation}.
The third-rank tensor was also used for bent-core liquid crystal molecules and is defined as
$
T_{ijk} 
=
\sum_{m=1}^{4}
\overline{
n_i^{(m)} n_j^{(m)} n_k^{(m)} 
}
,
$
where $n^{(m)}$ shows a vector of four axis pointing to vertices of a
tetrahedron \cite{Fel:1995}.
The third-rank tensor is  traceless in the sense that $T_{iik} = T_{ijj} = T_{iji} =
0$. 
Here we consider a slightly different definition of tensor order
parameters, namely we define $m$-rank tenors through the integral for
deformation $\delta R$
\begin{align}
\frac{R_0}{\Omega}
\int da
\overline{
n_{i_1} (a) n_{i_2} (a)
\cdots n_{i_m} 
}
\delta R(a)
\end{align}
(see Appendix~\ref{section.tensor.order.para}).
Here $\overline{n_{i_1} (a) n_{i_2} (a) \cdots n_{i_m}}$ denotes
traceless tensor constructed by $m$ unit normal vectors.
The tensors are consistent with those used in previous studies except normalization prefactors in the sense that both of
them have the
properties discussed below. 

\begin{figure}[h]
\begin{center}
\includegraphics[width=0.50\textwidth]{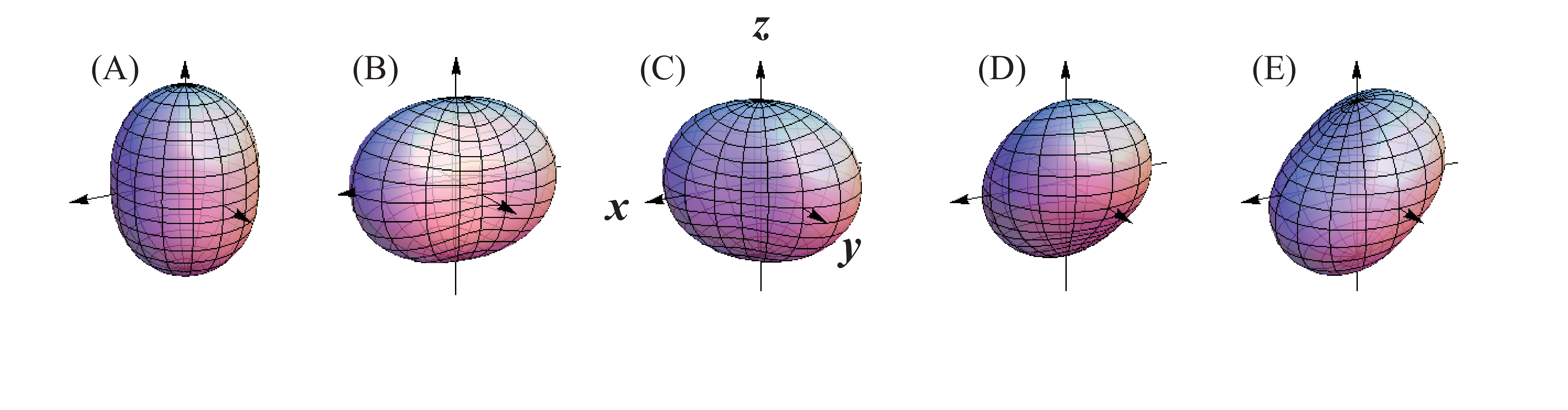}
\caption{ (Color Online)
Five independent shapes of a second mode ($l=2$).
Elongation in two directions ((A) and (B)) and shear deformation in
 three directions ((C)-(E)) are shown.
Each shape is expressed in terms of spherical harmonics as
(A) $w_{2,0}$, (B) $w_{2,2} + w_{2,-2}$, (C) $i(w_{2,2} - w_{2,-2})$,
(D) $w_{2,1}-w_{2,-1}$, and (E) $i(w_{2,1} + w_{2,-1}).$
\label{fig.n2shape}
}
\end{center}
\end{figure}

\begin{figure}[h]
\begin{center}
\includegraphics[width=0.50\textwidth]{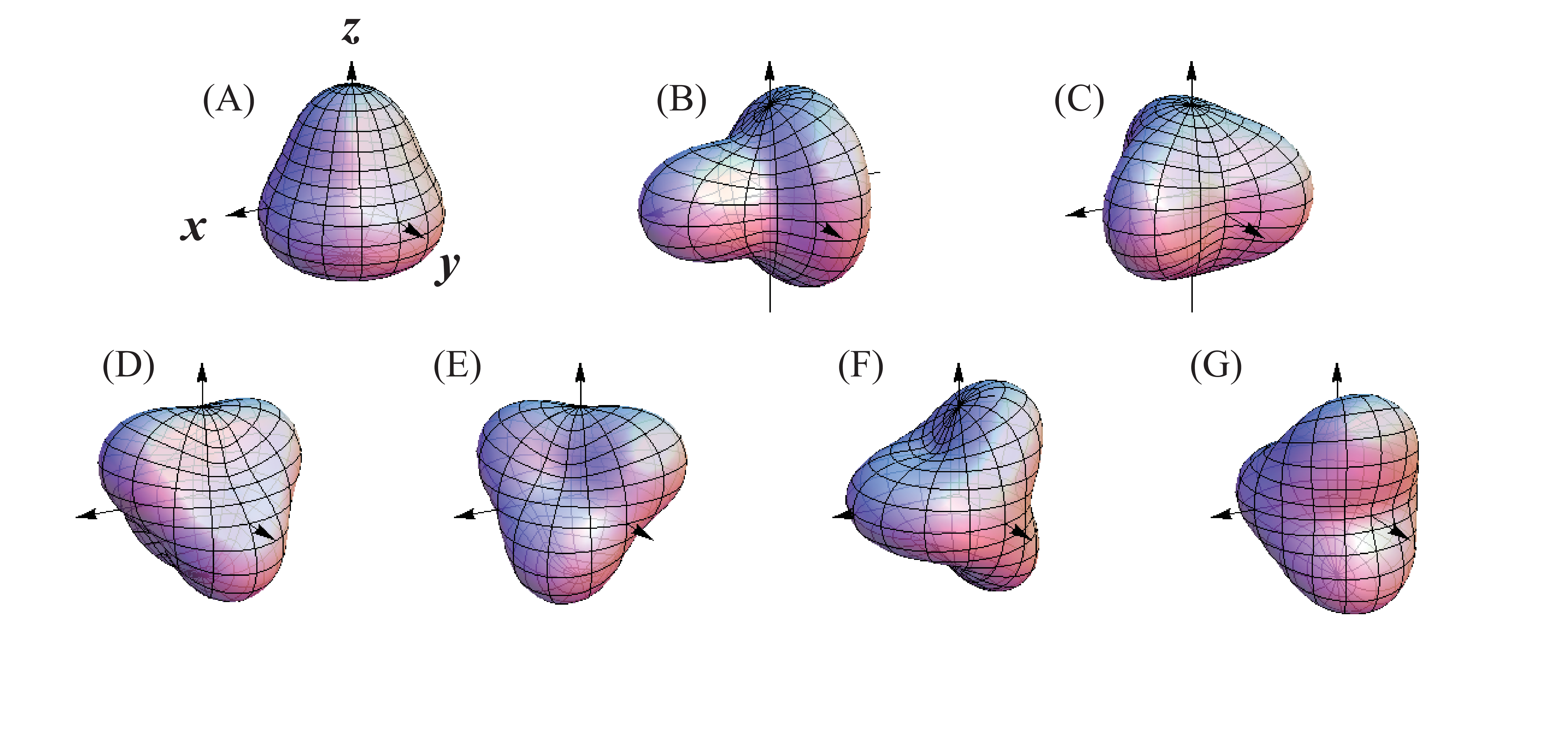}
\caption{ (Color Online)
Seven independent shapes of a third mode ($l=3$).
Each shape is expressed in terms of spherical harmonics as
(A) $w_{3,0}$, (B) $w_{3,-3}-w_{3,3}$, (C) $i(w_{3,-3}+w_{3,3})$,
(D) $w_{3,2}+w_{3,-2}$, (E) $i(w_{3,-2}-w_{3,2})$, (F) $w_{3,-1} -
 w_{3,1}$, (G) $i(w_{3,-1}+w_{3,1})$.
\label{fig.n3shape}
}
\end{center}
\end{figure}

The tensors are symmetric and traceless, and thus the number of
independent components is not the number of tensor elements.
The second-rank tensor $S_{ij}$ has five independent components, which
are given by the following irreducible forms \cite{Jerphagnon:1978}: $S_{33}$, $S_{11} - S_{22}$, $\frac{1}{2} (S_{12} + S_{21})$, $\frac{1}{2} (S_{13} +
S_{31})$, and $\frac{1}{2} (S_{23} + S_{32})$.
These correspond to two elongation and three shear deformation.
These independent are also expressed by the coefficients of spherical
harmonics as
$w_{2,0}$, $w_{2,2} + w_{2,-2}$, $i(w_{2,2} - w_{2,-2})$,
$w_{2,1}-w_{2,-1}$, and $i(w_{2,1} + w_{2,-1})$, respectively \cite{Jerphagnon:1978}.
The concrete shapes of these modes are described in
Fig. \ref{fig.n2shape}.
The third-rank tensor has seven independent components and therefore
there are seven irreducible forms \cite{Jerphagnon:1978}.
These are explicitly given as
\NY{
$2 T_{zzz} 
-  \left(
T_{xxz} + T_{xzx} + T_{zxx}
\right)
-
 \left(
T_{yyz} + T_{yzy} + T_{zyy}
\right)
$,
}
$T_{xxx} - T_{xyy}
- T_{yxy} - T_{yyx}
$,
$
T_{yyy} - T_{yxx}
- T_{xyx} - T_{xxy}
$,
$
\frac{1}{2}
\left(
T_{zxx} - T_{zyy}
\right)
+
\frac{1}{2}
\left(
T_{xzx} - T_{yzy}
\right)
+
\frac{1}{2}
\left(
T_{xxz} - T_{yyz}
\right)
$,
$
\frac{1}{6} \left(
T_{xyz} + T_{xzy} + T_{yxz}
+ T_{yzx} + T_{zxy} + T_{zyx}
\right)
$,
$
\frac{1}{3}
\left(
T_{xzz} + T_{zxz} + T_{zzx}
\right)
-\frac{1}{12}
\left(
T_{xyy} + T_{yxy} + T_{yyx}
\right)
- \frac{1}{4}
T_{xxx}
$, and
$
\frac{1}{3}
\left(
T_{yzz} + T_{zyz} + T_{zzy}
\right)
- \frac{1}{12}
\left(
T_{yxx} + T_{xyx} + T_{xxy}
\right)
- \frac{1}{4} T_{yyy}
$.
These independent components correspond to the coefficients of spherical
harmonics for $w_{3,0}$, $w_{3,-3}-w_{3,3}$,
$i(w_{3,-3}+w_{3,3})$, $w_{3,2}+w_{3,-2}$,
$i(w_{3,-2}-w_{3,2})$, $w_{3,-1} - w_{3,1}$, and
$i(w_{3,-1}+w_{3,1})$.
These shapes are shown in Fig.~\ref{fig.n3shape}.

In general, tensors of any rank appear.
Nevertheless each tensor order parameter of rank $m$ scales as $\mathcal{O}
(\epsilon^m)$ close to the critical point of drift bifurcation with a
small parameter $\epsilon \sim u$, droplet velocity as discussed in \cite{Shitara:2011}.
This is also demonstrated in Sec.~\ref{section.amplitude.eq}.
Therefore we consider expansion up to $\mathcal{O}(\epsilon^3)$, that is, a $l=2$ mode $S_{ij}$ and a $l=3$ mode $T_{ijk}$ (see Appendix~\ref{section.tensor.order.para}).

Once concentration is obtained, the kinematic condition
(\ref{vn.def.lin}) reads a time evolution equation for a shape of the droplet.
Once we obtain equations for $\dot{w}_{lm}$, we may obtain the equation for
a tensor order parameter
$\dot{S}_{ij}$ using the relation between spherical harmonics and tensor
order parameters (\cite{Shitara:2011,Fel:1995} and see also Appendix~\ref{section.tensor.order.para}).
The benefit of the above definition of tensor order parameters is that
we do not need to calculate kinematic condition for each $w_{lm}$, but instead
we constructed traceless tensor directly from deformation and the
kinematic condition is transformed into the equation of tensor order parameters.
The equation for a shape of the second mode is obtained by multiplying \NY{$
\frac{R_0}{\Omega}\int da \overline{n_i (a) n_j (a)}$} on
both sides of (\ref{vn.def.lin}) as 
\begin{align}
a_{2,0}^{(2)}
\odiff{S_{ij}}{t}
=& 
\frac{R_0}{\Omega}
\int da
\left[
v_n 
- ({\bf u} \cdot {\bf n}^{({\rm d})}) 
\right]
\overline{n_i (a) n_j (a)}
\nonumber \\
=&
\frac{R_0}{\Omega}
\int da
n_i (a) n_j (a)
v_n (a)
\nonumber \\
&
-
u_k 
\frac{R_0}{\Omega}
\int da
 n_k^{(d)} (a)
n_i (a) n_j (a)
\nonumber \\
&
- \frac{1}{3} \delta_{ij}
\frac{R_0}{\Omega}
\int da
\left[
v_n(a) - u_k n_k(a)
\right]
\nonumber \\
\simeq&
\frac{R_0}{\Omega}
\int da
\overline{n_i (a) n_j (a)}
v_n (a).
\label{eq.motion.n2}
\end{align}
Here we have used (\ref{intDYda}). 
We define a symmetric traceless tensor consisting of two vectors as
\begin{align}
 \overline{n_i (a) n_j (a)}
&=
n_i (a) n_j (a)
- \frac{1}{3} \delta_{ij}.
\end{align}
The term $-\frac{1}{3}\delta_{ij}$ is added in order to ensure the property of
traceless of the tensor order parameter.
Because of (\ref{unit.vec.def}), the second term
makes a contribution of $\mathcal{O}(u_k T_{ijk})$ and thus is neglected.
We rewrite (\ref{eq.motion.n2}) as
\begin{align}
a_{2,0}^{(2)}
\odiff{S_{ij}}{t} 
&=
M_{ij}^{(0)}+
M_{ij}^{(1)} + M_{ij}^{(2)},
\label{n2mode.decomp}
\end{align}
where $M_{ij}^{(0)}$ describes the relaxation of a shape from
(\ref{vn0}).
\NY{
\begin{align}
M_{ij}^{(0)} 
&=
- \frac{16 \gamma_0}{35 \eta R_0}
a_{2,0}^{(2)}
S_{ij}
=
-\kappa_{20} S_{ij}
.
\end{align}
}
$M_{ij}^{(1)}$
and $M_{ij}^{(2)}$ correspond to the
contribution from normal and tangential forces under an inhomogeneous
concentration field, respectively.
\NY{
\begin{align}
\begin{pmatrix}
M_{ij}^{(1)} \\
M_{ij}^{(2)} 
\end{pmatrix}
=&
\frac{R_0}{\Omega}
\int da 
n_i (a) n_j (a)
\begin{pmatrix}
v_{n,1} (a) \\
v_{n,2} (a)
\end{pmatrix}
\end{align}
where $v_{n,1} (a)$ and $v_{n,2} (a)$ have already been shown in
(\ref{vel.normal}), respectively.
}
At the linear order in deformation, these are calculated by the expansion of ${\bf n}$ with
(\ref{unit.vec.def}), $\kappa$ with (\ref{surface.tension.def.lin}),
$c({\bf r})$ with (\ref{cI}).
We decompose as $M_{ij}^{(1)}=\sum_{k=1}^4 M_{ij}^{(1,k)}$ and
$M_{ij}^{(2)} = \sum_{k=1}^4 M_{ij}^{(2,k)}$ 
\NY{
(see
Sec.~\ref{app.coeff} in \cite{noteSM} for the detail expressions).
}

In the same manner, the time evolution of a $m$-rank tensor is obtained
by multiplying a traceless tensor consisting of $\overline{ n_{i_1} n_{i_2}
n_{i_3} \cdots n_{i_m}}$.
Note that there is no coupling between higher modes in this framework
such as $S_{ij} T_{ijk}$.
This is because the velocity of the droplet $u_i$ is chosen as an
expansion parameter \NY{
close to the critical point of drift bifurcation.
}

A shape of the third mode, $T_{ijk}$, is obtained by multiplying 
$
 \frac{R_0}{\Omega}\int da
\overline{n_i (a) n_j (a) n_k (a)}
$
on
both sides of (\ref{vn.def.lin}).
Here we define the abbreviated notation of symmetric traceless tensor
consisting of unit normal vectors as
\begin{align}
&
\overline{n_i (a) n_j (a) n_k (a)}
\nonumber \\
 = &
n_i (a) n_j (a) n_k (a)  
-\frac{1}{5}
\left(
n_i (a) \delta_{jk}
 + n_j (a) \delta_{ik}
+ n_k (a) \delta_{ij}
\right).
\end{align}
We obtain
\begin{align}
a_{3,0}^{(3)}
\odiff{T_{ijk}}{t}
=& 
\frac{R_0}{\Omega}
\int da 
\left[
v_n 
- ({\bf u} \cdot {\bf n}^{({\rm d})}) 
\right]
\overline{n_i (a) n_j (a) n_k (a)}
\nonumber \\
\simeq&
\frac{R_0}{\Omega}
\int da
\overline{n_i (a) n_j (a) n_k (a)}
v_n (a)
\nonumber \\
&+
\frac{R_0}{\Omega}
\int da
\overline{n_i (a) n_j (a) n_k (a)}
\nabla_s \delta R(a)
.
\end{align}
The equation of motion is decomposed as 
\begin{align}
a_{3,0}^{(3)}
\odiff{T_{ijk}}{t}
=& 
N_{ijk}^{(0)} +
N_{ijk}^{(1)} + N_{ijk}^{(2)}
+ N_{ijk}^{(3)}
\label{n3mode.decomp}
\end{align}
where $N_{ijk}^{(0)}$ describes the relaxation of a shape from
(\ref{vn0}),
\NY{
\begin{align}
N_{ijk}^{(0)} 
&=
- \frac{16 \gamma_0}{21 \eta R_0}
a_{3,0}^{(2)}
T_{ijk}
=
- \kappa_{30} T_{ijk}
,
\end{align}
and
\begin{align}
\begin{pmatrix}
N_{ijk}^{(1)} \\
N_{ijk}^{(2)}
\end{pmatrix}  
&=
\frac{R_0}{\Omega}
\int da 
\overline{n_i (a) n_j (a) n_k (a)}
\begin{pmatrix}
v_{n,1} (a)\\
v_{n,2} (a)
\end{pmatrix}  
\end{align}
where $v_{n,1} (a)$ and $v_{n,2} (a)$ have already been shown in
(\ref{vel.normal}), respectively.
}
We may also have $N_{ijk}^{(3)}
=
-\frac{2}{7} \frac{a_{2,0}^{(2)}}{R_0} \overline{S_{ij} u_k}$.
We again decompose (\ref{n3mode.decomp}) as $N_{ijk}^{(1)}=\sum_{k=1}^4 N_{ijk}^{(1,k)}$ and
$N_{ijk}^{(2)} = \sum_{k=1}^4 N_{ijk}^{(2,k)}$ 
\NY{
(see
Sec.~\ref{app.coeff} in \cite{noteSM} for the detail expressions).
We expand these coefficients for $c=c^{(0)} + c^{(1)} + c^{(2)}+c^{(3)}$
and $\kappa$ in terms of $u_i$, $S_{ij}$, and $T_{ijk}$.
}


\section{amplitude equation}
\label{section.amplitude.eq}

\NY{
Combining the results in the previous sections, we obtain the following
amplitude equations for velocity $u_i$, deformation of second mode
$S_{ij}$, and deformation of third mode $T_{ijk}$ from (\ref{velocity.decomp}),
(\ref{n2mode.decomp}), and (\ref{n3mode.decomp}):
\begin{align}
m \odiff{u_i}{t}
 &=
(-1 + \tau) u_i
- g u^2 u_i
+ bu_j S_{ij}
,
\label{amp.eq.n1}
\\
\odiff{S_{ij}}{t} 
&=
- \left(
\kappa_2 + \kappa_{20} 
\right)
S_{ij}
+ \lambda 
\overline{u_i u_j}
\label{amp.eq.n2}
\\
\odiff{T_{ijk}}{t} 
=&
-\left(
\kappa_3 + \kappa_{30} 
\right)
T_{ijk}
+ \lambda_3
\overline{u_i u_j u_k}
+ b_3 
\overline{S_{jk} u_i}
\label{amp.eq.n3}
\end{align}
}
where the third-rank symmetric traceless tensor consisting of
second-rank tensor and vector is denoted as
\begin{align}
\overline{S_{jk} u_i}
 =&
\left[
\left(
S_{jk} u_i
+ S_{ik} u_j
+ S_{ij} u_k
\right)
\right.
\nonumber \\
&
\left.
- \frac{2}{5} u_{\delta}
\left(
S_{j \delta} \delta_{ik}
+ S_{k \delta} \delta_{ij}
+ S_{i\delta} \delta_{jk}
\right)
\right].
\end{align}
Here we focus on the system close to steady states and neglected the
terms of $\ddot{u}_i$, $u \dot{u}$, and $\dot{u}_j S_{ij}$.
Without deformation 
\NY{
that is, if we neglect (\ref{amp.eq.n2}) and (\ref{amp.eq.n3}), and set $S_{ij}=0$
, (\ref{amp.eq.n1}) recovers the results in \cite{Yabunaka:2012}
with the same coefficients.
The coefficients in (\ref{amp.eq.n1}) scale as 
\begin{align}
 \tau \sim \tau^* =
\frac{\gamma_c A}{D^2 \eta \beta^3}
\end{align}
and 
$
m \sim m^*
= \tau^* /(D \beta^2)
$, and $
g \sim g^*
=
\tau^*/(D^2 \beta^2)
$
(see Fig.1-3 in \cite{Yabunaka:2012})
.
The non-dimensional number $\tau^*$ characterize the activity of the system;
as the source term increases, spontaneous motion becomes more likely.
This is also dependent on the strength of chemo-mechanical coupling,
that is, as the flow field is more sensitive to the
surface tension, the activity is essentially enhanced.
In fact, the translational velocity of the droplet makes a transition from zero
to a finite value when $\tau^*>1$ as clearly seen in (\ref{amp.eq.n1}).
It is always the case that $\tau^* \geq 0$  and
therefore as the increase of the source
term proportional to $A$ and/or sensitivity of a surface tension to a
concentration field, the stationary state becomes unstable \cite{Yabunaka:2012}. 
}

The obtained equations for a shape have the same forms in the order
parameters ($u_i$,
$S_{ij}$, $T_{ijk}$) for the second mode in three dimensions
\cite{Shitara:2011,Hiraiwa:2011} and the third mode in two dimensions \cite{ohta:2009b}
if we use two-dimensional irreducible forms \footnote{We drop the term
$u_k T_{ijk}$ because this is higher $\mathcal{O} (\epsilon^4)$ order.}. 
However since we have seen the relation of a shape to
mechanical and flow properties in the previous sections, it is worth to
see the explicit forms of the coefficients.
The coefficients associated with deformation are expressed as
\NY{
\begin{align}
\kappa_2
&=
- 
\sum_{\beta=1}^4
\left(
M_{ij}^{(1,\beta,0)}
+ M_{ij}^{(2,\beta,0)}
\right)
= \kappa_2^* g_{\kappa_2} (x)
\\
\lambda
&=
M_{ij}^{(1,1,2)}
+ M_{ij}^{(2,1,2)}
=
- \lambda^* g_{\lambda} (x)
\label{lambda.Sij}
\end{align}
using the result of (\ref{int.nTnn.da1}) and (\ref{int.nTnn.da2}).
Here $g_{\kappa_2}(x)$ and $g_{\lambda}(x)$ are the functions of
$x=\beta R_0$ and the scaling of the coefficients $\lambda$, $\kappa_2$,
and $b$ in (\ref{coeff.b}) is
\begin{align}
 \lambda^*
& =
 \frac{4 \gamma_c A }{175 D^3 \eta \beta^4 a_{2,0}^{(2)}}
\sim
 \frac{ \gamma_c A }{D^3 \eta \beta^4}
=
\frac{\tau^*}{ D \beta}
.
\label{amp.coeff.scaling.lambda}
\\
\kappa_2^*
&=
 \frac{2 \gamma_c A R_0}{35 D \eta}
\sim
 \frac{\gamma_c A R_0}{ D \eta}
=
\tau^* D \beta^3 R_0
\\
b^*
&=
\frac{\gamma_c A}{D^2 \eta \beta^2}
=
\tau^* \beta
\label{amp.coeff.scaling.b}
.
\end{align}
The concrete forms of the functions $g_i(x)$ ($i=\kappa_2, \lambda_2,b$) are shown in Appendix.~\ref{app.coeff.Q}.

\begin{figure}[h]
\begin{center}
\includegraphics[width=0.50\textwidth]{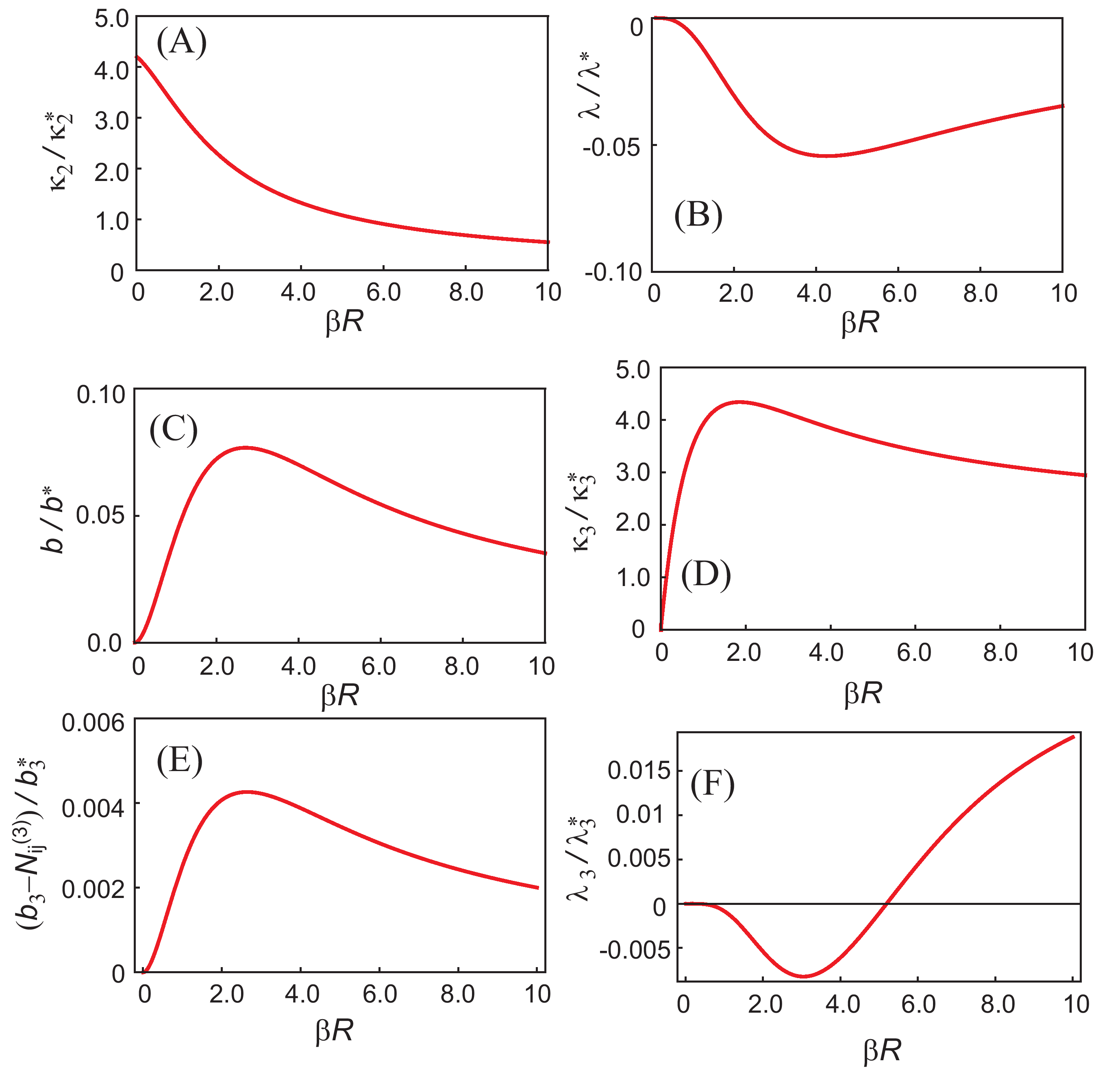}
\caption{(Color Online)
The coefficients in the amplitude equations of the second
 mode (\ref{amp.eq.n2}) and the third mode of deformation (\ref{amp.eq.n3}).
The values are normalized by (\ref{amp.coeff.scaling.lambda})-
 (\ref{amp.coeff.scaling.b}) and (\ref{amp.coeff.scaling.kappa3})-(\ref{amp.coeff.scaling.b3}).
\label{fig.coeff}
}
\end{center}
\end{figure}

The stationary shape is obtained from (\ref{amp.eq.n2})
\begin{align}
S_{ij} 
&=
\frac{\lambda}{\kappa_2 + \kappa_{20}}\left(
u_i u_j -\frac{1}{3} \delta_{ij}
\right).
\end{align}
}
As we have seen in Sec.~\ref{section.pusher}, properties of pusher or
puller are associated with a shape of the droplet.
Therefore, signs of $\lambda$ and $\kappa_2$ give us information about a
shape for the second mode.
Since we do not discuss any instability for the second mode itself,
$\kappa_2$ should be positive.
Indeed, the result shows $\kappa_2$ is always positive in our system
(Fig.~\ref{fig.coeff}A).
\NY{
This does not depend on the value of the contribution from
an inhomogeneous concentration $\kappa_2 + \kappa_{20}$ compared
with the contribution of the shape relaxation $\kappa_{20}$, which is
proportional to $\gamma_0$ and is always positive.
When the droplet moves parallel to the $z$-axis, the velocity is given
as $u_i = (0,0,u)$, and
\begin{align}
w_{2,0} 
&=
\frac{\lambda}{3 (\kappa_2 + \kappa_{20})}
u^2
\label{w20}
.
\end{align}
It is clear that the shape deviates from a sphere as the velocity
increases.
The shape is characterized by the coefficients $\lambda/\kappa_2$.
A function $\lambda/\lambda^*$ is
plotted in Fig. \ref{fig.coeff}B.
The result shows that $\lambda/\lambda^* \leq 0$.
We may evaluate $\lambda$ for $x \gg 1$ as
$
 \lambda
\simeq
-3 \lambda^*/(8x)
,
$
and for $x \ll 1$, it becomes
$
 \lambda
\simeq
- \lambda^* x^4/35
.
$
}
The droplet spontaneously moves when $\gamma_c >0$ for $A>0$.
In this case, we found it is always satisfied that $\lambda < 0$ (Fig. \ref{fig.coeff}B) and
accordingly $w_{2,0}<0$.
Therefore, the shape is elongated perpendicular to the moving direction.
This means that the droplet is characterized as a pusher.

\NY{
From (\ref{w20}), we have another non-dimensional number for isotropic surface
tension 
\begin{align}
\gamma^* 
&=
\frac{\gamma_0}{ \eta D \beta^3 R_0^2 }
\label{gamma0.nor}
\end{align}
characterizing time scale of deformation.
With this, deformation is rewritten with normalized velocity $\tilde{u}
= u/(D \beta)$ as 
\begin{align}
\beta  w_{2,0} 
& \simeq
 \frac{\tau^* \tilde{u}^2}{x (\tau^* + \gamma^*) }
\end{align}
showing that when $\tau^* \ll \gamma^*$, the droplet is almost sphere
and as $\tau^*$ increases it deforms proportional to $\tau^*/\gamma^*$.
This ratio characterizes the deformation of the self-propelled droplet
\begin{align}
\frac{\tau^*}{\gamma^*} 
&=
\frac{\gamma_c A R_0^2}{\gamma_0 D}
\label{ratio.t.g}
.
\end{align}
As the relaxation time scale of a concentration field ($\sim R_0^2/D$)
increases, that is the size of the droplet increases, the moving droplet is more likely
to deform.
}

\begin{figure}[h]
\begin{center}
\includegraphics[width=0.40\textwidth]{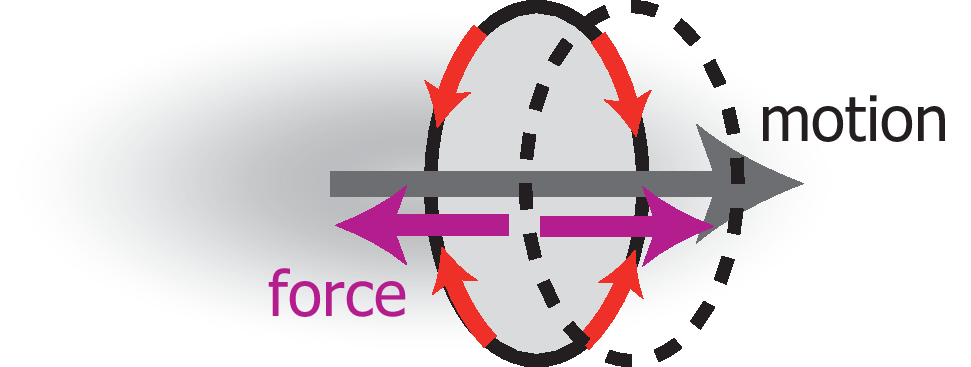}
\caption{(Color Online) 
Schematic illustration of a self-propelled droplet and a concentration
 field of chemical molecules that are produced by the droplet.
\label{fig.mechanism}
}
\end{center}
\end{figure}

The intuitive explanation of the sign of $\lambda$ is following.
When a droplet is moving in one direction, the concentration of
chemicals is higher behind the droplet and at the center in the axis
perpendicular to the direction of motion.
Therefore, the concentration is relatively higher at the front and the back, and
lower at the sides looking like a contrail.
The concentration distribution is schematically illustrated in Fig.~\ref{fig.mechanism}. 
For $\gamma_c>0$, the surface tension is also higher at the front and
the back, and
lower at the sides leading to deformation perpendicular to the direction of
motion. 
This mechanism also applies to the case for $\gamma_c <0$ and $A<0$
\cite{Yoshinaga:2012a} in which the concentration is lower at the front
and the back, and
higher at the sides.

We may also calculate another coefficient as
\begin{align}
b 
= &
\sum_{\beta=0}^5
u_{i,1}^{(\beta,1)}
+
\sum_{\beta=0}^4
u_{i,2}^{(\beta,1)}
=
b^* g_b (x)
\label{coeff.b}
\end{align}
If $b \lambda>0$, there is a Lyapunov function \cite{Shitara:2011}.
We found that $b$ is always positive and therefore there is no Lyapunov
function in this system.
This is not surprising in a sense that our model contains chemical
reaction and therefore is under nonequilibrium state.
However, given that a Lyapunov function does exist without deformation, it
may suggest the importance of coupling between velocity and shape.  

As seen in (\ref{amp.eq.n3}), the third mode is determined by the first
and second modes.
Therefore, the third mode is slaved by the other two modes. 
The coefficients arising from the third mode is
\NY{
\begin{align}
\kappa_3 
=&
- \sum_{\beta=1}^4
\left(
N_{ijk}^{(1,\beta,0)}
+ N_{ijk}^{(2,\beta,0)}
\right)
=
\kappa_3^* g_{\kappa_3} (x)
\label{amp.coeff.scaling.kappa3}
\nonumber \\
\lambda_3 
=&
N_{ijk}^{(1,1,3)} + N_{ijk}^{(1,1,3)}
= 
- \lambda_3^* g_{\lambda_3} (x)
\\
b_3 
&=
\sum_{\beta=1}^4
\left(
N_{ijk}^{(1,\beta,1)}
+ N_{i,j,k}^{(2,\beta,1)}
\right)
+ N_{ijk}^{(3)}
=
b_3^* g_{b_3}(x)
\label{amp.coeff.scaling.b3}
,
\end{align}
where $g_i(x)$ ($i=\kappa_3, \lambda_3,b_3$) are the functions of
$x=\beta R_0$ shown in Appendix.~\ref{app.coeff.Q}.
These coefficients are plotted in Fig.~\ref{fig.coeff} with following
normalization constants,
\begin{align}
\kappa_3^* 
&=
\frac{4 A \gamma_c}{105 \eta  D}
a_{3,0}^{(3)}
\sim
\frac{A \gamma_c}{\eta  D \beta}
= \tau^* D \beta^2
\\
\lambda_3^* 
&=
\frac{8 \gamma_c A}{1225 \eta  D^4}
\sim
\frac{\gamma_c A}{\eta  D^4 \beta^5}
=
\frac{\tau^*}{D^2 \beta^2}
\\
b_3^{*} 
&=
\frac{\gamma_c A}{\eta D^2} a_{2,0}^{(2)}
\sim
\frac{\gamma_c A }{\eta D^2 \beta^2}
=
\tau^* \beta
.
\end{align}
The relaxation of a shape is determined by $\kappa_3 + \kappa_{30}$, which is always
positive (Fig.~\ref{fig.coeff}D).
}
Therefore, there is no instability in this mode.


\subsection{inhomogeneous concentration distribution}
Similar to the shape and the velocity of the droplet, the concentration
field is also described in terms of tensors.
\NY{
The tensors are obtained by expanding the concentration field into
moments with $1$, $\overline{
n_i^{(0)} (\theta,\varphi)
n_j^{(0)} (\theta,\varphi)
}
$, $\overline{
n_i^{(0)} (\theta,\varphi)
n_j^{(0)} (\theta,\varphi)
n_k^{(0)} (\theta,\varphi)
}
$, and so on.
In particular, the second moment is given as
\begin{align}
C_{ij}^{(2)} (r)
&=
\frac{R_0}{\Omega}
\int 
\overline{
n_i^{(0)} (\theta,\varphi)
n_j^{(0)} (\theta,\varphi)
}
c({\bf r}) r^2 \sinT d\theta d\varphi.
\label{con.mode2}
\end{align}
According to (\ref{cI}), we may expand each mode of concentration fields for instance as $C^{(2)}_i =
C^{(2,0)}_i + C^{(2,1)}_i + C^{(2,2)}_i + C^{(2,3)}_i + \cdots$.
}
We are in particular interested in the second mode
\begin{align}
C_{ij}^{(2,0)} (R_0) 
&=
\frac{A}{D}
\bar{Q}_1^{(1)}
a_{2,0}^{(2)}
S_{ij}
\label{Cij20}
\end{align}
and
\NY{
\begin{align}
C_{ij}^{(2,2)} (R_0)
&\simeq
\frac{2A}{5D^3}
g_{\lambda} (x)
\left(
u_i u_j - \frac{1}{3} \delta_{ij}
\right)
.
\end{align}
}
Note that there is no contribution from $c^{(1)}$ on the second mode.
$C_{ij}^{(2,0)}$ describes the distortion of a concentration field due to
deformation of a droplet.
Suppose an elongated droplet in $z$-direction, that is, $S_{zz}>0$. 
Since $\bar{Q}_1^{(1)} > 0$, $C_{ij}^{(2,0)}$ is positive when the reaction
is $A>0$, that is creation inside the droplet. 
It
 shows an inhomogeneity of higher concentration along the $z$-axis
and lower in the $xy$-plane.
This results in higher surface tension along $z$-axis when $\gamma_c >0$
leading to relaxation toward a spherical shape is made.
On the contrary, $C_{ij}^{(2,2)}$ showing the distortion of concentration
due to the self-propulsive motion makes the droplet deformed.
In fact, $C_{ij}^{(2,2)}$ makes a higher concentration (higher surface tension) along the $z$-axis
and lower concentration (lower surface tension) in the $xy$-plane as we
discussed in (\ref{lambda.Sij}) for $\gamma_c A >0$.
The shape becomes elongated perpendicular to the $z$-axis due to
inhomogeneous surface tension, which is $\gamma_2 > 0$.
This argument also supports the system is characterized as a pusher.

\NY{
\subsection{helical motion}

Another point in the amplitude equations
(\ref{amp.eq.n1})-(\ref{amp.eq.n3}) is the time scales of motion and
deformation.
If the time scale of deformation is much faster than that of motion, the
dynamics of shape is slaved by the motion.
Therefore shape does not play a relevant role of the dynamics and we
expect straight motion.
When the time scale of deformation is comparable with that of motion,
the coupling between motion and shape affects on the dynamics.
In \cite{Hiraiwa:2011}, various types of motion have been discussed for
(\ref{amp.eq.n1})-(\ref{amp.eq.n3}) with arbitrary coefficients.
They found straight motion becomes unstable and helical motion appears
for some range of parameters.
Our model is based on mechanics and hydrodynamics, and the coefficients
are not independent.
Thus, it is of interest to see when the helical motion appears with our
physical coefficients.
In our notation, the helical motion is stable when \cite{Hiraiwa:2011}
\begin{align}
\frac{-1+\tau^*}{\tau^*} 
& \geq
\left(
\tau^* + \gamma^*
\right) x^2
+ x (\tau^* + \gamma^*)
\label{helical}
\end{align}
where $x=\beta R_0$.
From (\ref{helical}), we found that the helical motion is stable when $x
\ll 1$.
The time scale of velocity is $m/(-1+\tau) \sim (D \beta^2)^{-1}$ for
(\ref{amp.eq.n1}) while
the time scale of deformation is $(\kappa_2 + \kappa_{20})^{-1} \sim (D
\beta^3 R_0)^{-1}$ for (\ref{amp.eq.n2}).
The latter is more sensitive to $x=\beta R_0$ and thus as $x$ becomes
smaller, the time scale of deformation becomes slower. 
Note that for an arbitrary concentration distribution there is no screw
motion of the droplet along a straight path driven by a surface
tension gradient as (\ref{vel.3D.asym.exp2a}) suggests.
Nevertheless, the helical motion is realized by changing the path due to the coupling between
motion and deformation.
}

\subsection{active stress}

The inhomogeneous concentration is associated with stress acting on a
fluid through surface tension.
In this section, we discuss how the stress tensor is modified by
spontaneous motion and deformation.
The stress tensor arising from the inhomogeneous concentration is given
as \cite{Finlayson:1969,Yabunaka:2012}
\begin{align}
\sigma_{a} ({\bf r})
&=
- B_1 c({\bf r})
\nabla_i \phi \nabla_j \phi 
+ \mbox{isotropic terms}
\end{align}
where $\phi ({\bf r})$ is a phase specifying a droplet and surrounding
fluid.
The isotropic terms are absorbed into pressure because of incompressibility.
In the sharp interface limit, the stress is described by using a normal vector \cite{Yabunaka:2012}
\begin{align}
\sigma_{a} (a)
&=
\frac{1}{R_0} \int  \sigma_a ({\bf r}) dr
=
- \frac{\gamma_c}{R_0} c({\bf r})
n_i (a) n_j (a) .
\end{align}
Here the stress tensor is integrated over radial direction and therefore
defined on the surface.
In fact, this stress accumulates at the surface of the droplet.
\NY{
Integrating over the surface, we obtain
the anisotropic part of stress as 
\begin{align}
\sigma_{a} 
&\simeq
\sigma_a^*
\left(
u_i u_j
- \frac{1}{3} u^2 \delta_{ij} 
\right)
\\
\sigma_{a}^*
&=
\frac{\eta \lambda}{R_0} 
\simeq
- \frac{\gamma_c A}{D^2 \beta^3 R_0}
\end{align}
which characterizes active stress by $\sigma_a^*$.
}
Since $\lambda<0$, the active stress is negative in the direction of
motion.
This also suggests that the droplet is extensile, that is pusher.
If we regard the velocity $u_i$ as a polar vector $p_i$, then this
active stress is the same form used in active polar nematic liquid crystals \cite{kruse:2005}.

If we put the numbers as $D \sim 10^{-3}$  mm$^2$/s,  the velocity being $u
\sim D \beta$ from our results of $g$ and $\tau$, $R_0 \sim
100$ um, $\beta \sim R_0^{-1}$, and the characteristic change of surface
tension $\sim $ 1 mN/m with $\gamma_c A \sim $ 1 mN/(m
$\cdot$sec), we obtain $\sigma_a \sim 100$Pa.
The active stress estimated for lamellipodium is $\sigma_a \sim 1000$Pa
\cite{Juicher2007} and same order for actin cortex \cite{Salbreux2009}.

\section{Discussions}
\label{sec.discussions}

Our result is consistent with the phenomenological models proposed in
\cite{ohta:2009b}
where the amplitude equations are derived from
reaction-diffusion equations \cite{ohta:2009,Shitara:2011}.
Our approaches
including hydrodynamics has an advantage in that we stand on a
mechanical point of view and indeed we successfully relate a shape,
motion, and flow with force moments acting on the droplet.
\NY{
In fact, we obtain non-dimensional number (\ref{gamma0.nor}) and (\ref{ratio.t.g})
that control the deformation of the self-propelled droplet. 
The quantity could be measured and controlled in experiments.
We also note that our approach derives the amplitude equations
for tensor order parameters $u_i$, $S_{ij}$, and $T_{ijk}$ not by
calculating all the coefficients of spherical harmonics 
as in \cite{Shitara:2011}
, but rather
by translating the kinematic equation directly into the time evolution equations
of tensor order parameters .
The both approaches must lead to the same results but we believe that our
approach is easier and more systematic.
}

Without deformation, it has recently been reported in numerical simulations that a self-propelled droplet
has a property of pusher \cite{Schmitt:2013,Michelin:2012}.
Self-propulsion of a localized spot in a reaction-diffusion system also
shows deformation.
In \cite{Shitara:2011,ohta:2009}, it has also been shown
that a shape of the spot is
elongated perpendicular to the direction of motion.
They also obtained $\lambda <0$ and $b>0$ in our nomenclature (see
(\ref{lambda.Sij}) and  (\ref{coeff.b})).
In \cite{Kitahata:2013}, the motion of a camphor particle under a given
shape has been investigated.
The result again shows the direction of motion is perpendicular to the
long axis of an ellipsoidal shape, which is consistent with our result.
The model in \cite{Kitahata:2013} predicts $b<0$ for (\ref{coeff.b}) in our model, which leads to smaller
critical value of $\tau$ for the transition between stationary and
self-propelled states.
This is however different from our result of $b>0$.
Our model does not fix a shape and therefore the critical point does not
depend on a shape; the droplet is always spherical at the stationary
state while deformation occurs only when $\tau$ reaches at the critical
value.   
As we discussed in Sec.~\ref{section.amplitude.eq}, there is no Lyapunov
function in this system.
Therefore it is not possible to interpret a translation between $b$ and
$\lambda$ using a simple energetic argument.
Although there is no explicit interpretation of mechanical force in these
models, it is interesting that both the results of reaction-diffusion
equations and our model including hydrodynamics show the same shape of a spontaneously moving
object, namely elongated perpendicular to the direction.  
Further investigation concerning relations between reaction-diffusion models and
hydrodynamic models is left for future work.

\NY{
Our results show that a self-propelled droplet driven by chemical
reaction has a force dipole and it is characterized as pusher.
One may interested in how this conclusion depends on the model
(\ref{con1}).
Here we discuss how the results are dependent on the form of the source term.
First, it is stressed that the spontaneous motion occurs only for
$\gamma_c A >0$ and under this condition, the sign of $\gamma_c$ and $A$
does not matter.
The source term in (\ref{con1}) modifies $S_q$ in (\ref{Sq}), which is
dependent only on $\beta$ and $R_0$.
Therefore it does not modify the scaling of the coefficients $\tau^*$,$
m^*$, $g^*$, $\kappa^*$, $\lambda^*$, $b^*$, $\kappa_3^*$,
$\lambda_3^*$, and $b_3^*$ other than $\beta$ and $R_0$.
In our model, the source term has both an isotropic ($S_q^{(0)}$) and an
anisotropic ($S_q^{(1)}$) terms.
The latter arises from deformation.
If molecules are produced or consumed irrespective to a shape of
the droplet, the source term is independent from deformation and
$\bar{Q}_n^{(1)}$ in $g_i (x)$ obtained from the anisotropic term disappears.
Even in this case, $\lambda$ does not change and $\kappa$ is
not qualitatively modified, resulting in the same sign of a force dipole
and deformation.
From these arguments, although it is not conclusive for an arbitrary
function of the source term, we expect our results are applied in a wide range
of situations.
}

\section{Summary}
\label{sec.summary}

In this paper, we derive a set of equations for motion and deformation
of a self-propelled droplet driven by production of chemical molecules
from inside.
We interpret an inhomogeneous surface tension as a force acting on the interface of
the droplet and obtain the force moments acting on a surrounding fluid.
The force moments drive a flow and accordingly generate the spontaneous motion
and deformation of the droplet.
The motion of the droplet modifies a concentration field because of
production of molecules.
This feedback loop among position and shape of the droplet, flow field, and
concentration field sustains not only a steady motion but also deformation. 

We concentrate in this paper on a single droplet.
Interaction between deformed droplets is an interesting extension of this
work.
Due to the anisotropic character of each droplet, interaction may also
become anisotropic. 
We plan to explore these in future publications \cite{Note1}.

\begin{acknowledgments}
The author acknowledges Takao Ohta for bringing this problem to my attention.
The author is also grateful to Shunsuke Yabunaka, Tetsuya Hiraiwa, and
 Tanniemola Liverpool for
 helpful discussions and to the support by a Grant-in-Aid for Young Scientists
(B) (No.23740317).
\end{acknowledgments}

\appendix

\section{flow field in two dimensions}
\label{app.2D}
In two dimensions, the surface tension is expanded as
\begin{align}
\gamma (\theta)
&=
\sum_{n=0}^{\infty} \gamma_n \cos (n\theta)
\label{surface.tension.2D}
\end{align}
\NY{
and the force is expressed as the first line of (\ref{force.interface}).
Here we choose $x$-direction as a direction of motion.
The velocity field is expanded as (\ref{multipolev.3D}) 
with the Oseen tensor $\mathsf{T}_{ij}$ in two dimensions 
\begin{align}
\mathsf{T}_{ij}  
&=
 \frac{1}{4\pi \eta}
\left[
- \left(
\ln r 
\right)
\delta_{ij}
+ \frac{x_i x_j}{r^2} 
\right].
\end{align}
The first term in (\ref{multipolev.3D}) vanishes because there is no
force monopole $F_j^{(1)} 
=  0$ and the second term 
leads to the velocity field
}
\begin{align}
v_{r,l=2}
&=
 \frac{ \gamma_2 }{4\eta} \frac{R}{r} \cos 2\theta
\end{align}
and $v_{\theta,l=2}=0$, which is consistent with the velocity field obtained
from a boundary-value problem.
Note that the normal velocity obtained from the first moment is canceled
by a higher order term, and no deformation occurs in this case.
The third term in multipole expansion leads to the normal and tangential velocity field 
\NY{
\begin{align}
\left(
v_{r,l=1},
v_{\theta,l=1}
\right)
&=
\left(
u \frac{R^2}{r^2} \cosT
,
u \frac{R^2}{r^2} \sinT
\right)
\end{align}
where the velocity of the droplet is
$
u 
=
- \gamma_1 /(8  \eta).
$
As three dimensions, the velocity field for a force dipole decays slower
($\sim 1/r$) that that for quadrupole ($\sim 1/r^2$) and therefore
dominates interaction between swimmers.
}

\section{spherical harmonics and vector spherical harmonics}
\label{app.spherical.harmonics}

In order to avoid confusion about normalization, we list the properties
of spherical harmonics and vector spherical harmonics.
Spherical harmonics $Y_l^m (\theta,\varphi)$ are defined as
\begin{align}
Y_l^m (\theta,\varphi) 
&=
\sqrt{\frac{(2l+1)(l-m)!}{4\pi (l+m)!}} 
P_l^{m} (\cosT)e^{im\varphi}
,
\end{align}
with associated Legendre polynomial $P_l^{m}(\cosT)$ of degree $l$ and
order $m$ \cite{Arfken:1968}.

\NY{
Vector spherical harmonics are defined as \cite{Hill:1954}
\begin{align}
\begin{pmatrix}
{\bf Y}_l^m(\theta,\varphi) 
\\
{\bf \Psi}_l^m(\theta,\varphi) 
\\
{\bf \Phi}_l^m(\theta,\varphi) 
\end{pmatrix}
&=
\begin{pmatrix}
Y_l^m(\theta,\varphi) {\bf e}_r
\\
\pdiff{Y_l^m(\theta,\varphi) }{\theta} 
{\bf e}_{\theta}
+ 
\frac{1}{\sinT}
\pdiff{Y_l^m(\theta,\varphi) }{\varphi} 
{\bf e}_{\varphi}
\\
- \frac{1}{\sinT}
\pdiff{Y_l^m(\theta,\varphi) }{\varphi} 
{\bf e}_{\theta}
+ 
\pdiff{Y_l^m(\theta,\varphi) }{\theta} 
{\bf e}_{\varphi}
\end{pmatrix}
\end{align}
The orthogonal relation of vector spherical harmonics reads
$
{\bf Y}_l^m
\cdot {\bf \Psi}_{l'}^{m'} 
=
{\bf Y}_l^m
\cdot {\bf \Phi}_{l'}^{m'} 
=
{\bf \Psi}_l^m
\cdot {\bf \Phi}_l^m 
=0.
$
}
Note that ${\bf Y}_l^m$ and other two harmonics (${\bf \Psi}_{l'}^{m'}$
and ${\bf \Phi}_{l'}^{m'}$) are orthogonal irrespective to $l$, $l'$,
$m$, and $m'$ while ${\bf \Psi}_l^m$ and ${\bf \Phi}_l^m$ are orthogonal
only for $l=l$ and $m=m'$.

\section{Tensor order parameter}
\label{section.tensor.order.para}

In this appendix, we summarize tensor order parameters characterizing a
shape of a deformed droplet.
First, we consider the following integral
\begin{align}
&
\sum_{l,m}
\int da
n_i Y_l^{m*} w^*_{lm}
\nonumber \\
 = & 
\sqrt{\frac{4\pi}{3}}
\left(\frac{
- w^*_{1,1} + w^*_{1,-1}}
{\sqrt{2}},
i
 \frac{w^*_{1,1} + w^*_{1,-1}}
{\sqrt{2}},
w^*_{1,0}
\right)=0
\label{int.nYlm.da}
.
\end{align}
This integral vanishes because $w_{1m}$ corresponds to translational
motion and does not contribute to deformation
Nevertheless, it forms a tensor representation of a first mode expressed
by spherical harmonics for $l=1$.
Other
examples of this representation are found in (\ref{normal.vector}) and (\ref{multipole.drop.velocity}).

\subsection{second-rank tensor}

The second-rank tensor $S_{ij}$ is given by
\begin{align}
\frac{R_0}{\Omega}
\sum_{l,m}
\int da
\overline{
n_i (a) n_j(a) 
}
Y_l^{m}(a) c_{lm}
&=
\frac{1}{2\sqrt{5\pi}}
S_{ij}
=a_{2,0}^{(2)} S_{ij}
.
\label{int.nnY.da}
\end{align}
The product of two spherical harmonics is reduced to single spherical
harmonics using the Wigner 3j symbols \cite{Arfken:1968}.
Using this, the product of two normal vectors is expressed with spherical harmonics for $l=2$.
Then the tensor order parameter of the second is given as
\begin{align*}
S_{11} 
&=
\sqrt{\frac{3}{2}}
(w_{2,2} + w_{2,-2})
- w_{2,0}
\nonumber \\
&=
\sqrt{\frac{3}{2}}
(w_{2,2}^* + w_{2,-2}^*)
- w_{2,0}^*
\\
S_{12}
= S_{21} 
&=
i \sqrt{\frac{3}{2}}
(w_{2,2} - w_{2,-2})
=
i \sqrt{\frac{3}{2}}
(w_{2,-2}^* - w_{2,2}^*)
\\
S_{13}
=S_{31} 
&=
\sqrt{\frac{3}{2}}
(-w_{2,1} + w_{2,-1})
=
\sqrt{\frac{3}{2}}
(w_{2,-1}^* - w_{2,1}^*)
\\
S_{22} 
&=
- \sqrt{\frac{3}{2}}
(w_{2,2} + w_{2,-2})
- w_{2,0}
\nonumber \\
&=
- \sqrt{\frac{3}{2}}
(w_{2,2}^* + w_{2,-2}^*)
- w_{2,0}^*
\\
S_{23}
= S_{32} 
&=
i \sqrt{\frac{3}{2}}
(- w_{2,1} - w_{2,-1})
=
i \sqrt{\frac{3}{2}}
( w_{2,1}^* + w_{2,-1}^*)
\\
S_{33}
&=
-(S_{11} + S_{22}) =2 w_{2,0}
= 2 w_{2,0}^*
.
\end{align*}
Here we have used
$
Y_l^{m*} (\theta,\varphi) 
=
(-1)^m Y_l^{-m} (\theta,\varphi)
$.
The same form was obtained in \cite{Shitara:2011} but there is a small
difference arising from the different definition of spherical harmonics.

\subsection{third-rank tensor}

The third-rank tensor $T_{ijk}$ may be expressed using spherical harmonics, similar to
the second-rank tensor, as
\begin{align}
\frac{R_0}{\Omega}
\sum_{l,m}
\int da
n_i(a) n_j(a) 
n_k(a) 
Y_l^{m}(a) w_{lm}
&=
a_{3,0}^{(3)}
T_{ijk}
,
\label{int.nnnY.da}
\end{align}
where \NY{
$
 a_{3,0}^{(3)}
=
\frac{1}{5}
\sqrt{\frac{3}{7\pi}}
$
}
and each element is given as
\begin{align*}
&
T_{111} 
=
\frac{1}{4}
\left[
\sqrt{15} w_{3,-3}
- \sqrt{15} w_{3,3}
+ 3w_{3,1}
-3w_{3,-1}
\right]
\\
&
T_{222}
=
\frac{i }{4}
\left[
\sqrt{15} w_{3,-3}
+ \sqrt{15} w_{3,3}
+3w_{3,1}
+3w_{3,-1}
\right]
\\
&
T_{333}
=
\sqrt{3}
 w_{3,0}
\\
&
T_{112}=T_{121}=T_{211}
\nonumber \\
&= 
- \frac{i}{4}
\left[
\sqrt{15} w_{3,-3}
+ \sqrt{15} w_{3,3}
-w_{3,1}
-w_{3,-1}
\right]
\\
&
T_{113}=T_{131}=T_{311}
\nonumber \\
&=
\frac{1}{4 \sqrt{3}}
\left[
\sqrt{30} w_{3,-2}
+ \sqrt{30} w_{3,2}
- 6 w_{3,0}
\right]
\\
&
T_{223}=T_{232}=T_{322}
\nonumber \\
&=
-\frac{1}{4 \sqrt{3}}
\left[
\sqrt{30} w_{3,-2}
+ \sqrt{30} w_{3,2}
+ 6 w_{3,0}
\right]
\\
&
T_{123}=T_{312}=T_{231}
=T_{132}=T_{213}=T_{321}
\nonumber \\
&=
-\frac{i}{2}
\sqrt{\frac{5}{2}} 
\left(
w_{3,-2} - w_{3,2}
\right)
\\
&
T_{122}=T_{212}=T_{221}
\nonumber \\
&=
\frac{1}{4}
\left[
- \sqrt{15} w_{3,-3}
+ \sqrt{15} w_{3,3}
- w_{3,-1} + w_{3,1}
\right]
\\
&
T_{133}=T_{313}=T_{331}
=
 w_{3,-1} - w_{3,1}
\\
&
T_{233}=T_{323}=T_{332}
=
-i 
\left(
 w_{3,-1} + w_{3,1}
\right)
.
\end{align*}

\subsection{fourth-rank tensor}

The fourth-rank tensor $D_{ij \alpha \beta}$ is defined as.
\begin{align}
&
\frac{R_0}{\Omega}
\sum_{l,m}
\int da
n_i(a) n_j(a) 
n_{\alpha}(a) 
n_{\beta} (a)
Y_l^{m}(a) w_{lm}
\nonumber \\
= &
a_{4,0}^{(2)}
S_{ij\alpha\beta}
+
a_{4,0}^{(4)}
D_{ij\alpha\beta}
.
\label{int.nnnY.da}
\end{align}
where
\begin{align}
&
S_{ij\alpha\beta}
\nonumber \\
=& 
\frac{1}{6}
 \left(
S_{ij} \delta_{\alpha \beta}
+ S_{\alpha \beta} \delta_{ij}
+ S_{i\alpha} \delta_{j \beta}
+ S_{i\beta} \delta_{j  \alpha}
+ S_{j\alpha} \delta_{i \beta}
+ S_{j\beta} \delta_{i \alpha}
\right)
\end{align}
and
\NY{
$
a_{4,0}^{(2)} 
=
\frac{6}{7} a_{2,0}^{(2)}
$
}
.
The fourth-rank tensor has symmetry such that
$
D_{ij\alpha\beta} 
= D_{ji\alpha \beta}
= D_{ij \beta \alpha},
$
and therefore the tensor is expressed as a $6\times 6$ matrix (Voigt representation).
In addition, the tensors $S_{ij\alpha\beta}$ and $D_{ij\alpha\beta}$ are
symmetric also in the representation,
$
 D_{ij\alpha\beta} 
=
D_{\alpha\beta ij}. 
$

\section{properties of unit normal vector and spherical harmonics}
In this section, we summarize useful properties of unit normal vectors
on a sphere and spherical harmonics.
First, it is readily shown 
\begin{align}
 \int da n_i(a)
&=0.
\label{int.norb.da}
\end{align}
The product of the normal vector is integrated as \cite{ohta:2001}
\begin{align}
\frac{R}{\Omega} \int da 
n_i(a) n_j(a)
&=
\delta_{ij}
\label{intnnda}
,
\\
 \frac{R}{\Omega} \int da 
n_i(a) n_j(a) n_k(a) n_l(a)
&=
\frac{1}{5}
\left(
\delta_{ij} \delta_{kl}
+ \delta_{ik}\delta_{jl}
+ \delta_{il} \delta_{jk}
\right).\label{intnnnnda}
\end{align}
In order to calculate the third mode, we also need
\begin{align}
&
 \frac{R_0}{\Omega} 
\int da 
n_i(a) n_j(a) n_k(a) 
n_{\alpha}(a) n_{\beta}(a)
n_{\gamma}(a)
= 
\frac{1}{35}
[\![
\delta_{ij} \delta_{k \alpha} \delta_{\beta \gamma}
]\!]
\end{align}
where $[\![]\!]$ implies a sum of 14 other permutations among all indices.

Next, we consider the integral including the derivative of spherical harmonics.
First, we have
\begin{align}
\frac{R_0}{\Omega}
\int da  
\nabla_{s,i} Y_l^m (\theta,\varphi) 
w_{lm}
&=0
\label{intDYda}
\end{align}
since $w_{lm}=0$ for $l=1$.
Using integral by part, the integral including a normal vector and
surface gradient operator acting on spherical harmonics is expressed as
\begin{align}
&
\frac{R_0}{\Omega}
\sum_{l,m}
\int da n_i 
\nabla_{s,j} Y_l^m (\theta,\varphi) 
\nonumber \\
=&
\frac{1}{\Omega}
\sum_{l,m}
\int da n_i 
\left[
t_j
 \pdiff{}{\theta}
+
b_j
\frac{1}{\sinT} \pdiff{}{\varphi}
\right]
Y_l^m (\theta,\varphi) 
\nonumber \\
= &
a_{1,1}^{(2)} S_{ij}
\label{intnDYda}
\end{align}
\NY{
with
$
a_{1,1}^{(2)}
=
\frac{3}{R_0} a_{2,0}^{(2)}
$.
In the same manner, we may list the following calculations
\begin{align}
&
\frac{R_0}{\Omega}
\sum_{l,m}
\int da 
\begin{Bmatrix}
n_i(a)  n_j (a) 
\nabla_{s,k} 
 \\
n_i (a) n_j (a) n_k (a) 
\nabla_{s,\alpha}
\\
\nabla_{s,i} \nabla_{s,j} 
\\
n_i (a)
\nabla_{s,\alpha} \nabla_{s,\beta}
\\
n_i (a) n_j (a)
\nabla_{s,\alpha} \nabla_{s,\beta} 
\end{Bmatrix}
Y_l^m (\theta,\varphi) 
w_{lm} 
\nonumber \\
= &
\begin{Bmatrix}
a_{2,1}^{(3)} T_{ijk}
\\
a_{3,1}^{(2)} S_{ijk\alpha}^{(3,1)} 
+
a_{3,1}^{(4)} D_{ijk\alpha} 
\\
a_{0,2}^{(2)} S_{ij}
\\
a_{1,2}^{(3)} T_{ijk}
\\
a_{2,2}^{(2)} S_{ij\alpha \beta}^{(2,2)}
+  a_{2,2}^{(4)} D_{ij \alpha \beta}
\end{Bmatrix}
\label{intnDYda2}
\end{align}
where 
$
 a_{2,1}^{(3)}
=
4 a_{3,0}^{(3)}
$,
$
a_{3,1}^{(2)} 
= 
\frac{1}{7 R_0} a_{2,0}^{(2)}
$, 
$
 a_{0,2}^{(2)}
= \frac{6}{R_0^2}
a_{2,0}^{(2)}
$, 
$
 a_{1,2}^{(3)}
=
12 a_{3,0}^{(3)}
$, and $a_{2,2}^{(2)} 
=
\frac{1}{7 R_0^2} a_{2,0}^{(2)}
$.
The fourth-rank tensors in these formula are
\begin{align}
S_{ijk\alpha}^{(3,1)}
= &
-2 
\left(
S_{ij} \delta_{k \alpha}
+ S_{i k} \delta_{j \alpha}
+ S_{j k} \delta_{i \alpha}
\right)
\nonumber \\
&
+ 5 
\left(
S_{k \alpha } \delta_{ij}
+ S_{j \alpha} \delta_{i k}
+ S_{i \alpha} \delta_{j k}
\right)
\end{align}
and
\begin{align}
S_{ij\alpha \beta}^{(2,2)}
=&
-8 S_{ij} \delta_{\alpha \beta}
+ 20 S_{\alpha \beta} \delta_{ij}
-8 S_{i \alpha} \delta_{j \beta}
\nonumber \\
&
- S_{j \beta} \delta_{i \alpha}
- S_{i \beta} \delta_{j \alpha}
-8 S_{j \alpha} \delta_{i \beta}
.
\end{align}
}


\section{The functions $g_i (x)$}
\label{app.coeff.Q}

In this section, we summarize the function $g_i (x)$ in the coefficients of
the amplitude equations.
In the second mode,
\begin{align}
g_{\kappa_2} (x)
&=
10 \frac{Q_1^{(0)}}{R_0}
-3
\pdiff{Q_1^{(0)}}{s}
+  \bar{Q}_{1,2}^{(1)}
\\
g_{\lambda}
&=
- \frac{1}{R_0} \pdiff{Q_3^{(0)}}{s}
+ \pdiff{^2 Q_3^{(0)}}{s^2}
\\
g_b (x)
=&
- 
\left[
\frac{1}{15}
\pdiff{\bar{Q}_{2,2}^{(1)}}{s}
+ \frac{2}{5}\frac{\bar{Q}_{2,2}^{(1)}}{R_0}
+ \frac{17}{15 R_0} 
\pdiff{Q_2^{(0)}}{s}
- \frac{3}{5} \pdiff{^2 Q_2^{(0)}}{s^2}
\right]
.
\end{align}
For third mode,
\begin{align}
 g_{\kappa_3}(x)
=&
 \bar{Q}_{1,3}^{(1)}
+ \frac{22}{R_0} Q_1^{(0)}
- 
3 \pdiff{Q_1^{(0)}}{s} 
\\
g_{\lambda_3} (x)
=&
\frac{3}{R_0^2} \pdiff{Q_4^{(0)}}{s}
- \frac{3}{R_0} \pdiff{^2 Q_4^{(0)}}{s^2}
+ \pdiff{^3 Q_4^{(0)}}{s^3}
\\
g_{b_3}(x)
=&
 \frac{\gamma_c A}{735 \eta D^2} 
\frac{a_{2,0}^{(2)}}{a_{3,0}^{(3)}}
\left[
\frac{8}{ R_0} \bar{Q}_{2,2}^{(1)}
-4
\pdiff{\bar{Q}_{2,2}^{(1)}}{s}
\right.
\nonumber \\
&
\left.
+
8 \pdiff{^2 {Q}_{2}^{(0)}}{s^2}
-
\frac{42}{ R_0} \pdiff{{Q}_{2}^{(0)}}{s}
\right]
-\frac{2}{7} \frac{a_{2,0}^{(2)}}{a_{3,0}^{(3)} R_0}
\end{align}


\begin{thebibliography}{60}
\bibitem{noteAvel}%
  \BibitemOpen
  \bibinfo {note} {{Strictly speaking, this velocity is velocity of the center
  of mass, which is defined by $ {\bf R}_A = \Omega^{-1} \int {\bf X} dV. $ The
  position vector is defined in an arbitrary polar coordinate $ {\bf X} = {\bf
  X}(r,\theta,\varphi). $ Assuming that the volume does not change $ (\partial
  / \partial t) \left( \Omega {\bf R}_A \right) = (\partial / \partial t) \int
  {\bf X} dV = \int {\bf X} v_n da. $ this is equivalent to
  (\ref{vel.geometric}). We may also define the contour center of a surface $ {\bf
  R}_L = A^{-1} \int {\bf X}(a) da, $ which in general different from ${\bf
  R}_A$ for large deformation. }}%
  \bibAnnoteFile{NoStop}{noteAvel}%
\end{thebibliography}



\end{document}